\documentclass[twocolumn]{aa}
\usepackage{graphicx}
\usepackage{txfonts}

\begin{document}

   \title{Nuclear activity in galaxy pairs: a spectroscopic analysis
   of 48 UZC-BGPs. \thanks{Based on observations obtained with the Cassini telescope in Loiano (BO)}}

   \author{ P.,\, Focardi \inst 1, V. \,Zitelli \inst 2, S. \, Marinoni \inst 1  \inst ,\inst 3}


   \institute{ Dipartimento di Astronomia, Universit\'a di Bologna, Italy\\
              \email{paola.focardi@unibo.it}
  \and
             INAF-OABO, Via Ranzani 1, 40127 Bologna, Italy
\and
             Fundacion Galileo Galilei \& TNG, PO Box 565 S/C de la Palma,
         Tenerife Spain \\}


   \date{Received ; Accepted }

   \abstract
 {The role played by interaction on galaxy formation and evolution
   is a long lasting debated subject.   Several questions remain still open, 
   among them if, and to what extent, galaxy interaction may induce nuclear activity
 }
   { Galaxy pairs are ideal sites in which to investigate the role of interaction on nuclear activity,
,for
this reason we have undertaken a spectroscopic survey of a large
homogeneous sample of galaxy pairs (UZC-BGP) 
}
 {We present the results of the nuclear spectral classification, 
  of 48 UZC-BGPs, which represent more than half of the whole sample and have an excellent
morphological match with it.}
   {The fraction of emission line galaxies 
    is extremely large, especially
   among spirals where it reaches 84 \% and 95 \%, for early and late spirals respectively.
   SB (Star Burst), is the most frequent type of nuclear activity encountered
   (30 \% of galaxies),
   while AGNs (Active Galactic Nuclei)
   are only 19\%. The fractions raise to  45 \% and 22 \% when considering only spirals.
   Late spirals are characterized by both an unusual increase (35 \%) of AGN activity
   and high luminosity (44 \% have  M$_B$  $<$   -20.0 + 5log $h $). LLAGNs (Low Luminosity AGNs)  are only 8\% of the
   total number of galaxies, but this kind of activity could be present in another 10 \% of the galaxies
   (LLAGN candidates). 
Absorption line galaxies reside mostly (61 \%) in S0 galaxies and display the lowest B luminosity in
    the sample, only 18 \% of them have  M$_B$  $<$ -20 + 5 log $h$, but together
with LLAGNs (candidates included) they are the most massive galaxies in the sample.
Intense-SB nuclei are found in galaxy pairs  with  galaxy-galaxy projected
separations up to 160  $h$$^{-1}$ kpc suggesting that in bright isolated galaxy pairs
interaction may be at work and effective up to that distance.}
   {AGNs are characterized by an advanced morphological type while SB phenomenon occurs with the same
   frequency in early and late spirals. 
LLAGNs and LLAGN candidates do not always show similar properties, the former
   are more luminous in B, richer in early-type (E-S0s) galaxies, and half of  them are hosted in galaxies
   showing visible signs of interaction with fainter companions.
   This last finding  suggests that minor interactions might be a driving mechanism
for a relevant fraction of LLAGNs. The differences between LLAGNs and LLAGN candidates might
confirm, instead,  the  heterogeneous nature of this class of objects.}


\authorrunning{Focardi et al.}
\titlerunning {Nuclear activity in UZC-BGPs}
   \maketitle
%

\section{Introduction}
The role played by interaction on galaxy formation and evolution is
a long lasting largely debated subject  involving both   the ``far''
and ``local'' universe. Hierarchical models of galaxy formation
invoke, in fact, large occurrence of interaction and merging
phenomena, which are expected to increase with redshift (Governato
et al. 1999; Gottlober et al. 2001) and affect morphologies, gas
distribution and population of galaxies (Dubinski et al. 1996; Mihos
\& Hernquist 1996). Evolution of the merging rate, with z, has been
estimated using close galaxy pairs, but results are, however,
conflicting  (Zepf \& Koo 1989; Carlberg et al. 1994; Le Fevre et
al. 2000; Bundy et al. 2004), which is not unexpected due to
differences in sample depths, observational techniques and selection
criteria (Patton et al. 2000). Interaction and mergers are less
frequent in  the ``local'' universe, but can be analysed with higher
detail. Nearby galaxy systems can be investigated at faint
luminosity, on a wide spatial scale and with a better knowledge of
surrounding environment. However, even  ``locally'' 
theoretical expectations
have not found adequate support from observational data.
Galaxy interaction   should be, in fact,
extremely effective in
 redistributing large amounts of material
  towards the galaxy central regions, giving raise, in this way, to violent bursts
   of star formation (\cite{barnes91}).
  Tidal interaction between galaxies is expected
   to induce instabilities in the discs, able to generate
  galaxy bar formations, which, producing
  an inflow of gas towards the galaxy central regions, may even activate AGN
  phenomenon (Noguchi 1988; Barnes \& Hernquist 1991).
  However, observations concerning the amount and level of nuclear activity
  in nearby interacting systems have produced so far conflicting results neither
  able to confirm, nor to definitely rule out theoretical expectations.
  In fact, even though, starting with
   Larson \& Tinsley (1978), there has been
 growing evidence
  of an increase of
 star formation in interacting galaxy systems (e.g., Kennicut \& Keel 1984;
 Kennicut et al. 1987; Keel 1993, 1996; Donzelli \& Pastoriza 1997; Barton et al.
 2000), a one-to-one correlation between galaxy-galaxy interaction
 and star formation remains unclear. Such a correlation holds only for an extremely
 limited number of objects (ULIRGs, Sanders \& Mirabel 1996) which show a fraction
 of interacting galaxies nearly close to 100 \% (Sanders et al. 1988; Borne et al.
 1999), while there are several interacting systems with no sign of star
 formation. The situation becomes even more complex and controversial for the so-called
 AGN-interaction paradigm for which conflicting results have been given so far (Dahari
 1985; Keel et al. 1985; Fuentes-Williams \& Stocke 1988; Rafanelli et al. 1995;
 MacKenty 1989; Kelm et al. 1998;  De Robertis et al. 1998; Schmitt 2001; Kelm
 et al. 2004). However, large part of this contradiction is likely to be ascribed to
 inhomogeneities among the analyzed samples which are often small,
 have
 been selected by different methods and criteria, and may be biased towards
 or against certain kind of systems.

Galaxy pairs are ideal sites to investigate the role of   interaction on nuclear activity
  since proximity in redshift and in
projected separation make interaction and encounters between member galaxies highly probable.
 The recent availability of a large and complete
 nearby 3D galaxy catalog (UZC, Falco et al.1999), has made  it feasible to select
 a volume-limited sample of 89
 Bright
 Galaxy Pairs  (UZC-BGPs, Focardi et al. 2006, hereafter paper I) which
 does not suffer from  velocity/distance biases or contamination by projection effects.
  At variance with previous available nearby pair samples  (KPG, Karachentsev et al.
  1972, RR Reduzzi \& Rampazzo 1989), the first
  selected visually
  from the POSS plates, the second applying KPG criteria to the ESO-LV catalog
  (Lauberts \& Valentijn 1989), the UZC-BGP sample has been selected
  by means of
  an objective neighbour
  search algorithm (Focardi \& Kelm 2002) applied to the UZC catalog;
  it is thus complete, homogeneous and contains pairs which are
 already close in 3D space.

 The analysis of UZC-BGP, based on available data (
 paper I), has allowed us to show that ellipticals are extremely rare
 and  underluminous (in B), while late spirals ($>$ Sc) are overluminous. This
 finding confirms previous
 claims (Kelm \& Focardi 2004) linking the
   formation of bright ellipticals
   to group/cluster environment
    and suggest that galaxy-galaxy interaction might be responsible
    of the blue luminosity enhancement
  of disk galaxies through SF phenomena. This last suggestion found
  support in the strong FIR emission displayed by
   a significant fraction of early spirals,
   mostly belonging to interacting pairs.

 Very few data are available (see paper I)
 concerning nuclear activity type in UZC-BGP;
 we have thus undertaken a
 spectroscopic survey of the sample.
 The survey is currently still
 ongoing (now 85 \% complete).
  In this paper we present results and analysis
 for 48 UZC-BGP which constitute more than half of the whole sample and
 have an excellent morphological match with it.

  The sample is presented in (\S 2); in
    (\S 3) we show the results of our  nuclear activity  classification based on standard
    diagnostic diagrams; in
 (\S 4) we analyze and compare the characteristics of galaxies
 having different nuclear activity type ; in (\S  5) we look for
 possible link of nuclear activity with interaction strength. The conclusions
 are drawn in \S 6.
 In analogy with paper I,  a Hubble constant
 of $H_0 = 100\, {\it h}$ \, km \, s$^{-1}$\, Mpc$^{-1}$
  is assumed throughout.


\section{The sample}


\begin{table*}
\begin{center}
\caption[] {The E+E pairs}
\begin{tabular}{||l|r|l|c|l||}
\hline
Name & UZC-BGP & Type & T   & Emission lines\\
\hline
\hline
 NGC 2991   & 23A & S0 & -2.1 & -- \\
  NGC 2994 &23B  & S0 & -2.0 & --   \\
  NGC 3567  &   35A  & S0 & -2.1 & --  \\
   CGCG 039-055  & 35B & E & -2.7 & --  \\
     CGCG 160-030  & 51A & S0/a & -1.4 & --\\
 CGCG 160-036  & 51B  & E & -4.3 &  [NII]$_{6583}$ \\
  CGCG 102-023  & 58A &  S0 & -1.7 &  -- \\
  CGCG 102-024  & 58B &  S0 & -1.8 & --  \\
   IC 999  &  63A &  S0 &-1.9 & -- \\
  IC 1000 &  63B & S0 & -2.0 & -- \\
   NGC 6018 & 77A & SB0/a  & -1.2 & [NII]$_{6548}$, H$_{\alpha}$,
   [NII]$_{6583}$,[SII]$_{6717}$,[SII]$_{6731}$\\
   NGC 6021 &  77B  & E & -4.0 & [NII]$_{6583}$ \\
 \hline
\end{tabular}
\end{center}
\end{table*}

The sample contains 48 galaxy pairs, which represent more than half (54 \%) of UZC-BGP  sample.\, The latter
  is a volume-limited sample
of galaxy pairs which has been selected from UZC catalog applying  an adapted version of
the neighbour search algorithm of Focardi \& Kelm (2002).
\, The environment of each UZC galaxy  having M$_{Zw}$ $\le$ -18.87
+ 5 log $h$,  v$_r$   $\in [2500
 $-$  7500]$\ km s$^{-1}$ and  $|$ b$^{II}$$|$
$\ge$ 30$^0$  has been explored on a surronding area characterized by two
projected dimensions (r$_p$ = 200  $h$ $^{-1}$ kpc and  R$_p$ = 1 $h$$^{-1}$ Mpc)  
and  a radial velocity ``distance''  ($|$ $\Delta$ v$_r$ $|$  = 1000 km s$^{-1}$).
Galaxies having to only one bright  neigbour ( M$_{Zw}$ $\le$ -18.87
+ 5 log $h$) within  r$_p$ and $|$ $\Delta$ v$_r$ $|$ and no other ones up to
  R$_p$  (and within $|$ $\Delta$ v$_r$ $|$) entered the UZC-BGP sample.
 The adopted value for  r$_p$ (galaxy-galaxy projected distance) accounts for possible huge haloes tied to bright
galaxies (Bahcall et al. 1995; Zaritsky et al. 1997). The 
value for  $|$ $\Delta$ v$_r$ $|$  is
large enough to not induce an artificial
cut in the relative velocities of galaxies in pairs (within  r$_p$) and to prevent contamination by
galaxy groups/clusters (up to  R$_p$). The value for R$_p$ (large scale
isolation radius) was chosen to ensure  the absence of luminous companions on group/cluster typical scale.       
  The lower limit in radial velocity was fixed to
reduce distance uncertainities due to peculiar motions and to prevent contamination
by the Virgo complex, while the upper limit was set to guarantee 
sampling of UZC luminosity function just below L$^*$ (Cuesta-Bolao \& Serna 2003).
Finally, the limit in $|$ b$^{II}$$|$ was imposed to minimize the effects of galactic
absorption. 

At variance with other galaxy pair samples selected from 3D catalogs (Barton et al. 2000; Alonso et al. 2004), 
which are magnitude limited and contain pairs belonging to any kind of large scale environment,  
UZC-BGP is particularly suited to investigate the mutual effect of two
bright close companions, isolated on the typical group/cluster scale. Minor companions,
which might have failed either UZC (m$_{Zw}$ $\le$ 15.5) or UZC-BGP (M$_{Zw}$ $\le$ -18.87
+ 5 log $h$) luminosity limit,
could be present in the local (within r$_p$) or distant (within R$_p$) environment but are not expected to play
a role comparable to the one of the two massive galaxies in the pair.
 UZC-BGP
galaxies are, in fact,  rather bright and, since luminosity relates to mass although not in an obvious way, 
rather massive \footnote{a rough estimate of the minimum mass ($M_{min}$) of UZC-BGP galaxies can be derived from their minimum luminosity  
L$_B$ $\sim$ 5.5 10$^9$ $h$$^{-2}$ L$_{\sun}$ and assuming, for them, an average $M$/L of 5 $h$$^{-1}$,
this gives  $M_{min}$ $\sim$  2.8 10$^{10}$ $h$$^{-1}$ $M_{\sun}$.}
 too. 
(Further details on UZC-BGP sample can be found in paper I). 

Following Karachentsev (1972) galaxy pairs can be classified on the
basis of their morphological content in E+E, S+S and E+S. E+E
contain only early-type galaxies (E + S0s) , S+S only spirals and
E+S  both types. The sample we present here is a fair representation
of the whole UZC-BGP as it contains 6 E+E (12 \%), 23 S+S (48 \%) and
19 E+S (40\%), to be compared with 13 \%, 48 \% and 39 \% respectively.

\begin{table*}
\begin{center}
\caption[] {The S+S pairs}
\begin{tabular}{||l|r|l|c|l||}
\hline
Name & UZC-BGP & Type & T & Emission lines\\
\hline
\hline
 NGC 23  & 2A & SBa & 1.2 & [OII]$_{3727}$, H$_{\beta}$,
 [OIII]$_{4959}$,[OIII]$_{5007}$, [OI]$_{6300}$, [NII]$_{6548}$, H$_{\alpha}$, [NII]$_{6583}$, [SII]$_{6717}$, [SII]$_{6731}$  \\
 NGC 26  & 2B & Sab & 2.4 & [NII]$_{6583}$ \\
 NGC 800  &  8A  & Sc & 5.3 & --  \\
 NGC 799  & 8B & SBa & 1.1 &  [NII]$_{6548}$, H$_{\alpha}$, [NII]$_{6548}$, [SII]$_{6717}$, [SII]$_{6731}$ \\
  NGC 871   &  9A & SBc & 4.6 & [OII]$_{3727}$, H$_{\beta}$,[OIII]$_{4959}$,[OIII]$_{5007}$, [OI]$_{6300}$,[NII]$_{6548}$, H$_{\alpha}$,
  [NII]$_{6583}$,[SII]$_{6717}$, [SII]$_{6731}$, HeI$_{7065}$  \\
  NGC 877   & 9B & SBc & 4.8 & [NII]$_{6548}$, H$_{\alpha}$, [NII]$_{6583}$, [SII]$_{6717}$, [SII]$_{6731}$   \\
  UGC 4074 &   17A & Sc  & 5.9 & H$_{\alpha}$ \\
  CGCG 262-048  & 17B & Sc & 5.0 & [NII]$_{6548}$, H$_{\alpha}$, [NII]$_{6583}$\\
   CGCG 035-023  & 22A & S? & 1.6& H$_{\beta}$, [OIII]$_{5007}$, [OI]$_{6300}$, [NII]$_{6548}$, H$_{\alpha}$, [NII]$_{6583}$, [SII]$_{6717}$, [SII]$_{6731}$   \\
  CGCG 035-25   & 22B & S? & 2.8  & H$_{\beta}$,[OIII]$_{5007}$, [NII]$_{6548}$, H$_{\alpha}$, [NII]$_{6583}$, [SII]$_{6717}$, [SII]$_{6731}$  \\
    UGC 5241         & 24A & Sbc & 4.4 & H$_{\beta}$, [OIII]$_{4959}$,[OIII]$_{5007}$,HeII$_{5412}$,[OI]$_{6300}$,H$_{\alpha}$, [NII]$_{6583}$, [SII]$_{6717}$, [SII]$_{6731}$   \\
    CGCG 265-043  & 24B &  Sbc & 4.3 & [NII]$_{6548}$, H$_{\alpha}$, [NII]$_{6583}$  \\
    NGC 3303 & 28A & Sb  &3.2 & [OII]$_{3727}$ , H$_{\beta}$,
    [OIII]$_{4959}$,[OIII]$_{5007}$, [OI]$_{6300}$,[NII]$_{6548}$, H$_{\alpha}$,[SII]$_{6717}$, [SII]$_{6731}$      \\
   CGCG 094-098  &  28B  & S? & 3.8 &  [OII]$_{3727}$, H$_{\beta}$, [OIII]$_{4959}$,[OIII]$_{5007}$, [OI]$_{6300}$,[NII]$_{6548}$, H$\alpha$, [NII]$_{6583}$,
    [SII]$_{6717}$, [SII]$_{6731}$     \\
    MKN 725 & 29A  &  Sc & 4.7&  H$_{\beta}$,[OIII]$_{4959}$, [OIII]$_{5007}$, H$_{\alpha}$, [NII]$_{6583}$,
   [SII]$_{6717}$, [SII]$_{6731}$, HeI$_{7065}$ \\
  UGC 5822  & 29B  & SBa &  1.0 & [OII]$_{3727}$, H$_{\gamma}$,H$_{\beta}$,
  [OIII]$_{4959}$,[OIII]$_{5007}$,HeI$_{5876}$, [OI]$_{6300}$, [NII]$_{6548}$, H$_{\alpha}$, 
[NII]$_{6583}$, [SII]$_{6717}$, [SII]$_{6731}$  \\
    UGC 6033    & 31A  &  Sb & 3.0 & --- \\
   CGCG 213-008  & 31B & S? & 0.7&  --- \\
  UGC 6397&  36A  &Sab & 2.0 & [NII]$_{6548}$, H$_{\alpha}$, [NII]$_{6583}$\\
  CGCG 185-053  &  36B & Sab &1.5 & [NII]$_{6548}$, H$_{\alpha}$, [NII]$_{6583}$,[SII]$_{6717}$, [SII]$_{6731}$ \\
 NGC 3719  & 37A    & Sbc & 3.7&   [NII]$_{6548}$, H$_{\alpha}$, [NII]$_{6583}$  \\
 NGC 3720 &  37B &  Sa & 1.4&  H$_{\alpha}$, [NII]$_{6583}$, [SII]$_{6717}$, [SII]$_{6731}$ \\
  UGC 7383  &   44A  & Sab &2.0 &  H$_{\beta}$,[OIII]$_{5007}$, H$_{\alpha}$, [NII]$_{6583}$ \\
   VCC 0395 &44B & Sbc & 4.5& [NII]$_{6548}$, H$_{\alpha}$, [NII]$_{6583}$  \\
    IC 962  & 59A & Sa &0.6  & [NII]$_{6548}$, H$_{\alpha}$, [NII]$_{6583}$\\
 CGCG 074-014  & 59B & Sab & 1.7& [OIII]$_{4959}$,[OIII]$_{5007}$, [NII]$_{6548}$, H$_{\alpha}$, [NII]$_{6583}$, [SII]$_{6717}$,[SII]$_{6731}$ \\
   Mrk 820  &66A &  S0/a & 0.1& [OII]$_{3727}$, H$_{\beta}$, [OIII]$_{4959}$,[OIII]$_{5007}$,[NII]$_{6548}$, H$_{\alpha}$, [NII]$_{6583}$, [SII]$_{6717}$, [SII]$_{6731}$   \\
  UGC 9464  & 66B  & S0/a & 0.0 & -- \\
  IC 1075   &68A &    SBb & 3.4& [OII]$_{3727}$, H$_{\beta}$, [OIII]$_{4959}$,[OIII]$_{5007}$, HeI$_{5896}$, [OI]$_{6300}$,[NII]$_{6548}$, H$_{\alpha}$, [NII]$_{6583}$, [SII]$_{6717}$, [SII]$_{6731}$  \\
  IC 1076   & 68B &   Sb & 2.8 & [NII]$_{6548}$, H$_{\alpha}$, [NII]$_{6583}$, [SII]$_{6717}$,[SII]$_{6731}$   \\
  NGC 5797 &69A &  S0/a & 0.0 & -- \\
  NGC 5804 &  69B & SBb & 3.1 &[OII]$_{3727}$, H$_{\gamma}$,H$_{\beta}$, [OIII]$_{4959}$,[OIII]$_{5007}$, [OI]$_{6300}$,[NII]$_{6548}$, H$_{\alpha}$, [NII]$_{6583}$, [SII]$_{6717}$, [SII]$_{6731}$    \\
  NGC 5857  &  71A &   SBab & 2.4 & [NII]$_{6583}$,\\
   NGC 5859 &71B &   SBbc & 4.0 & [NII]$_{6548}$, H$_{\alpha}$, [NII]$_{6583}$, [SII]$_{6717}$,[SII]$_{6731}$  \\
    CGCG 354-023   &   74A & S?  &2.8 & [OII]$_{3727}$, H$_{\beta}$, [OIII]$_{4959}$,[OIII]$_{5007}$, [OI]$_{6300}$,[NII]$_{6548}$, H$_{\alpha}$, [NII]$_{6583}$, [SII]$_{6717}$, [SII]$_{6731}$  \\
  CGCG 354-023  & 74B  & S? & 2.8 & [OII]$_{3727}$, H$_{\beta}$, [OIII]$_{4959}$,[OIII]$_{5007}$, [OI]$_{6300}$,[NII]$_{6548}$, H$_{\alpha}$, [NII]$_{6583}$, [SII]$_{6717}$, [SII]$_{6731}$   \\
   IC 4576  & 76A &  S?  & 0.3& --\\
 IC 4577    & 76B   & S? & 1.0& --\\
   NGC 6246 &   80A & SBb & 3.1 &   H$_{\alpha}$  \\
  NGC 6246 A    &  80B &  SBc & 5.1& [NII]$_{6548}$, H$_{\alpha}$, [NII]$_{6583}$, [SII]$_{6717}$,[SII]$_{6731}$     \\
  CGCG 452-021  & 81A &   Sc & 5.8& H$_{\beta}$, [OIII]$_{4959}$,[OIII]$_{5007}$,[NII]$_{6548}$, H$_{\alpha}$, [NII]$_{6583}$, [SII]$_{6717}$, [SII]$_{6731}$ \\
 UGC 12067  & 81B &   Sa & 1.0 & H$_{\beta}$, [OIII]$_{4959}$,[OIII]$_{5007}$, [OI]$_{6300}$,[NII]$_{6548}$, H$_{\alpha}$, [NII]$_{6583}$, [SII]$_{6717}$, [SII]$_{6731}$ \\
 IC 5242    &   82A&    Sa & 1.9 &[OII]$_{3727}$, H$_{\beta}$,[OIII]$_{5007}$,[NII]$_{6548}$, H$_{\alpha}$, [NII]$_{6583}$, [SII]$_{6583}$, [SII]$_{6731}$  \\
  IC 5243   &  82B   & Sbc & 3.6 &  H$_{\alpha}$  \\
 Mrk 308    &   83A & S0/a & 0.3 & [OII]$_{3727}$, H$_{\gamma}$,H$_{\beta}$ ,[OIII]$_{4959}$,[OIII]$_{5007}$,HeI$_{5876}$, [OI]$_{6300}$,
 [OI]$_{6364}$,[NII]$_{6548}$, H$_{\alpha}$, [NII]$_{6583}$,\\
& & &   & [SII]$_{6717}$, [SII]$_{6731}$ , HeI$_{7065}$, [OII]$_{7325}$ \\
 KUG 2239+200A  &   83B &  Sbc & 4.4 & [OII]$_{3727}$, H$_{\gamma}$, H$_{\beta}$ ,[OIII]$_{4959}$,[OIII]$_{5007}$,HeI$_{5876}$, [OI]$_{6300}$,[NII]$_{6548}$, H$_{\alpha}$, [NII]$_{6583}$,
 [SII]$_{6717}$,\\
& & & &  [SII]$_{6731}$ , HeI$_{7065}$ \\

    \hline
\end{tabular}
\end{center}
\end{table*}

Two dimensional long slit  spectra have been acquired with BFOSC
(the Bologna Faint Object Spectrograph and Camera) at the 152 cm
telescope (of Bologna University) in Loiano. Spectral coverage is
4000-8500 \, \AA \, with an average  resolution of about 4 \AA. The
slit was positioned on the nucleus of each galaxy and its width was
set either at 2" or at 2.5", (depending on seeing conditions), which
corresponds to galaxy nuclear region (about 500 $h^{-1}$  pc
at the average redshift of our sample).  The whole data
reduction was performed with IRAF. \,After standard CCD (flat field
and bias) correction, we extracted the spectra, calibrated them in
wavelength, identified the emission lines and measured their EW.
Spectral extraction was limited to the 4-5 central pixels,
corresponding to 2''-2.5''  at the detector scale. We set the
continuum level in the close neighborhood of each emission line (on
both sides) and when the emissions ( H$_\alpha$ and/or H$_\beta$)
were affected by the presence of an underlaying strong absorption
line, we set the continuum at the bottom of the emission line, to
minimize the effect of the absorption.

 Galaxies in E+E, S+S
and E+S pairs are listed respectively in Tables 1, 2 and 3. In each
Table we give  the galaxy  name, (column 1), the UZC-BGP
identificator (column 2), morphological classification (type) and
type code (T) (both from LEDA, columns 3 and 4), emission lines
identified in each spectrum, if any, (column 5). We have identified
only lines having
 S/N $\ge$ 5, very few lines, however, are characterized by such a low signal,
 the vast majority has, on average,
 S/N $\ge  50$. Morphology is very well defined for 86 \% (83/96) of the galaxies, less
defined for 13 galaxies, in these last cases the morphological classification  (column 3)
is followed by a question mark.

\begin{table*}
\begin{center}
\caption[] {The E+S pairs}
\begin{tabular}{||l|r|l|r|l||}
\hline
Name & UZC-BGP & Type & T & Emission lines\\
\hline
\hline
  NGC 41    &  3A &  Sc & 5.0& [OII]$_{3727}$, [NII]$_{6548}$, H$_{\alpha}$, [NII]$_{6583}$,[SII]$_{6717}$, [SII]$_{6731}$ \\
   NGC 42   & 3B & E/S0 &-3.0 & --\\
 NGC 160 & 4A & S0/a & -0.4 & --\\
 NGC 169 & 4B& Sab & 2.5 & H$_{\alpha}$,[NII]$_{6583}$, [SII]$_{6731}$\\
 NGC 192    &  5A &  SBa & 1.0 & H$_{\beta}$, [OIII]$_{4959}$,[OIII]$_{5007}$,[NII]$_{6548}$, H$\alpha$, [NII]$_{6583}$,[SII]$_{6717}$,
  [SII]$_{6731}$   \\
   NGC 196  & 5B &  SB0 & -1.8 &   --\\
 NGC 997S  &  10A &E & -3.8 & [NII]$_{6548}$, H$_{\alpha}$, [NII]$_{6583}$ \\
  NGC 998  &  10B &  S?& 2.4 & [NII]$_{6548}$, H$_{\alpha}$, [NII]$_{6583}$\\
  Mkn 1076 & 14A & Sbc & 3.8 & [OII]$_{3727}$, HeI$_{4026}$, H$_{\gamma}$, HeII$_{4686}$ H$_{\beta}$,
  [OIII]$_{4959}$,[OIII]$_{5007}$, [OI]$_{6300}$,[NII]$_{6548}$, H$_{\alpha}$,
  [NII]$_{6583}$, [SII]$_{6717}$,\\
  & & &   & [SII]$_{6731}$ \\
  CGCG 390-059  & 14B & E/S0 &-2.9 &--\\
  NGC 1587 & 16A & E & -4.8  & --  \\
 NGC 1589 &  16B & Sab  & 2.3 & [NII]$_{6583}$\\
  NGC 2528 &  18A &  SBb & 3.1 &  H$_{\gamma}$, H$_{\beta}$,[NII]$_{6548}$, H$_{\alpha}$,
  [NII]$_{6583}$,[SII]$_{6717}$, [SII]$_{6731}$ \\
 NGC 2524 & 18B &   S0/a & -0.2 & --\\
NGC 2744 &   20A &  SBab & 2.1&  [OII]$_{3727}$, HeI$_{4026}$, H$_{\delta}$, H$_{\gamma}$, H$_{\beta}$,
[OIII]$_{4959}$,[OIII]$_{5007}$, HeI$_{5876}$,[OI]$_{6300}$, [NII]$_{6548}$, H$_{\alpha}$, [NII]$_{6583}$, \\
& & &  & [SII]$_{6717}$, [SII]$_{6731}$\\
  NGC 2749 & 20B  & E & -4.8   & [NII]$_{6548}$, H$_{\alpha}$, [NII]$_{6583}$,[SII]$_{6717}$, [SII]$_{6731}$ \\
 NGC 2872 & 21A &  E & -4.8 & --\\
 NGC 2874 &  21B & SBbc &4.4 & [NII]$_{6548}$, H$_{\alpha}$, [NII]$_{6583}$,[SII]$_{6717}$, [SII]$_{6731}$\\
   CGCG 008-034  &  26A & Sb & 2.5 & [NII]$_{6583}$ \\
 IC 590 N   &   26B &  E/S0 & -3.2 & H$_{\alpha}$, [NII]$_{6583}$    \\
  CGCG 065-023  &27A & S0/a & -1.0 & [NII]$_{6548}$, H$_{\alpha}$, [NII]$_{6583}$ \\
  UGC 5627  & 27B &  Sab &  1.7 &  [NII]$_{6548}$, H$_{\alpha}$, [NII]$_{6583}$\\
  NGC 4003  &  39A &  SB0 & -1.9 & H$_{\alpha}$, [NII]$_{6583}$ \\
  NGC 4002  & 39B &   Sa  & 1.4& [NII]$_{6583}$\\
 NGC 4446   &  49A &  Sc & 5.8&  H$_{\beta}$, [OI]$_{6300}$,[NII]$_{6548}$, H$_{\alpha}$, [NII]$_{6583}$\\
  NGC 4447  &  49B & SB0 & -2.2 & --\\
 CGCG162-059 & 61A & SBbc & 4.4 & [NII]$_{6548}$, H$_{\alpha}$, [NII]$_{6583}$\\
 UGC 9012 & 61B & S0 & -2.0 & -- \\
  UGC 9413  & 65A & Sbc  & 4.0 & [OII]$_{3727}$,H$_{\beta}$,
[OIII]$_{4959}$,[OIII]$_{5007}$, [NII]$_{6548}$, H$_{\alpha}$,
[NII]$_{6583}$, HeI$_{6678}$ [SII]$_{6717}$, [SII]$_{6731}$  \\
   CGCG 353-044 &  65B & S0 & -1.7 & -- \\
  NGC 5771  &  67A & E & -4.5 & --\\
  NGC 5773  &  67B &  Sa & 1.4& -- \\
 CGCG 136-013  &  75A &  E? & -0.8 &  [NII]$_{6583}$\\
  CGCG 136-015  &  75B &  S? & 2.2  & [NII]$_{6548}$, H$_{\alpha}$, [NII]$_{6583}$ \\
 NGC 6251   & 79A &  E & -4.8  &  [OIII]$_{5007}$, [NII]$_{6548}$, H$_{\alpha}$, [NII]$_{6583}$ \\
   NGC 6252 & 79B &   S? & 4.4 &  [NII]$_{6583}$\\
  IC 5285   &  84A& E? & -0.8 & [NII]$_{6583}$\\
NGC 7489 & 84B &Sc & 6.5 & H$_{\alpha}$, [NII]$_{6583}$\\

  \hline
\end{tabular}
\end{center}
\end{table*}

Inspection of Tables 1 to 3 provides evidence that emission
line galaxies are extremely rare in E+E, over abundant in S+S  and rather
frequent (more than
half of the galaxies) in  E+S pairs. In fact, there are 3 galaxies with emission
 lines in the E+E pairs representing 25 \% of the total number of galaxies, 39 emission line galaxies in the
S+S (85 \%) and 26 in the E+S (68 \%) pairs. The different fractions are obviously related to the different
morphological content of the pairs and this is clearly illustrated in Fig. 1,
whose 4 panels
show the morphological distribution of galaxies in the whole sample
(upper left), E+E (upper right), S+S (lower left) and E+S (lower
right) pairs. Morphology is represented, on the x axis, by means of
the type code T (column 4 in each Table), which is a numerical
parametrization of the morphological type, introduced by de
Vaucouleurs. According to this parametrization, early-type galaxies
(E and S0s) have T $<$ 0 (E in general T $<$ -3), while late spirals
($>$ Sbc) have, on average, T $\ge 4$. The increase of emission line
galaxies with  morphology  is clearly evident in all pair samples
containing spiral galaxies (panels 1, 3  and 4 of Fig. 1) and
especially in panels 1 and 4 which evidence the larger
occurrence of emission features in spiral than in E-S0 galaxies.

The much higher frequency of emission line galaxies among spirals
than among early-type (E-S0s) galaxies  is not unexpected as
emission features occur more frequently in gas rich than in gas poor
galaxies; however the two lower panels of Fig.1 show that our sample
is characterized by a morphological content which is never more
advanced than T=6 (corresponding roughly to Sc galaxies). The
frequency of emission lines among spirals attains the maximum value
for morphologies more advanced than Sc as the the ratio of the
current SFR (Star Formation Rate) to the average past one increases
from about 0.01 in Sa to 1 in Sc-Irr galaxies (Kennicut et al.
1994).  The large frequency of emission lines occurring among early
spirals in our sample might thus be partly induced by interaction.

\begin{figure}
   \centering
  \includegraphics[width=8cm]{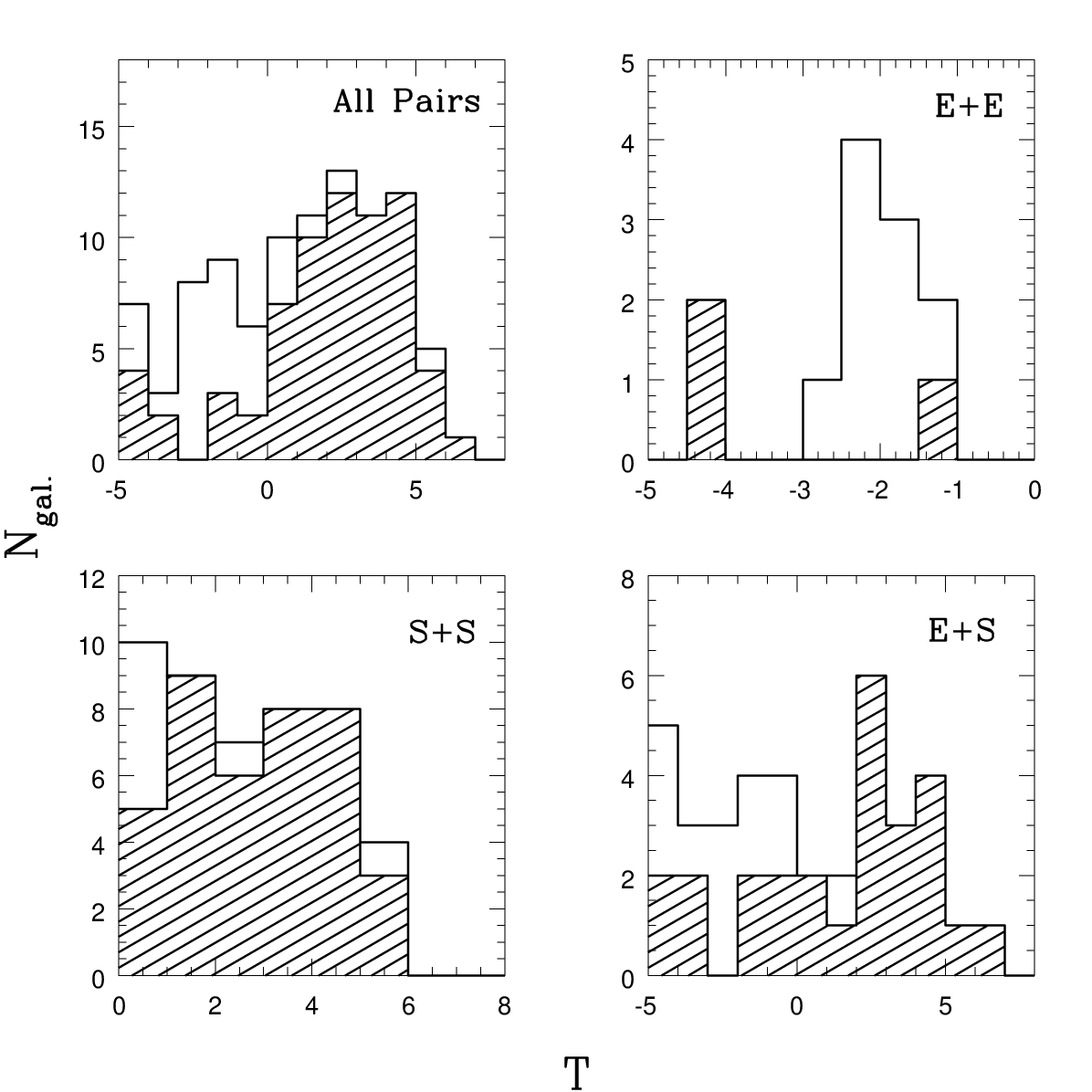}
  \caption{Relation between pair morphology  and the presence of
  emission lines in the member galaxy spectra.
  Continuous histogram show the morphological distribution of the whole
  galaxy pair sample (upper left), of E+E (upper right), S+S (lower left)
  and E+S (lower right) pairs. Dashed histograms indicate morphological
  distribution of galaxies with emission line spectrum, in each sample.}
    \end{figure}

The two low panels of Fig. 1 show clearly that the fraction of
emission galaxies  is much lower in E+S than in S+S pairs 
simply because of  the  large content of early-type (E-S0s)
galaxies in E+S pairs. In fact,  in this last sample,
 11 of the 12 galaxies
with only absorption lines in their spectrum are early-type
galaxies, implying
that the fractions of early-type galaxies and spirals with
emission lines are 42 \% (8/19) and 95 \% (18/19) respectively.
If we compare the last fraction  with the fraction (85 \%) of emission line
galaxies in S+S we see that both
E+S and S+S pairs
have an equally large probability of hosting an emission line
spiral.
    \begin{figure}
   \centering
  \includegraphics[width=8cm]{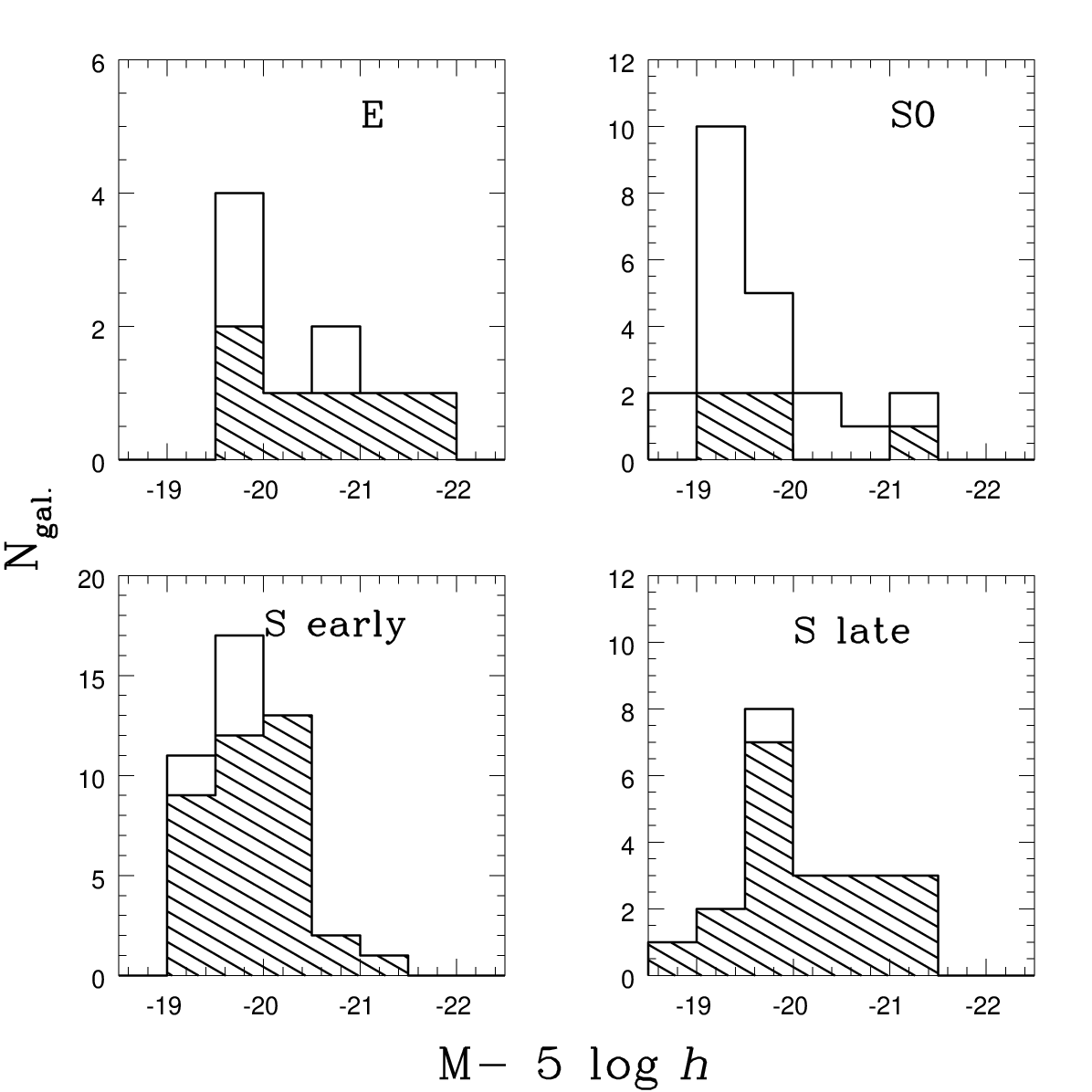}
  \caption{Absolute magnitude (M$_B$) distribution of  galaxies, of
  different morphological types, having or not having emission lines
  in their spectra. Continuous distribution refer to the whole samples,
  hatched one to galaxies with emission lines.}
    \end{figure}
    \begin{figure}
   \centering
  \includegraphics[width=8cm]{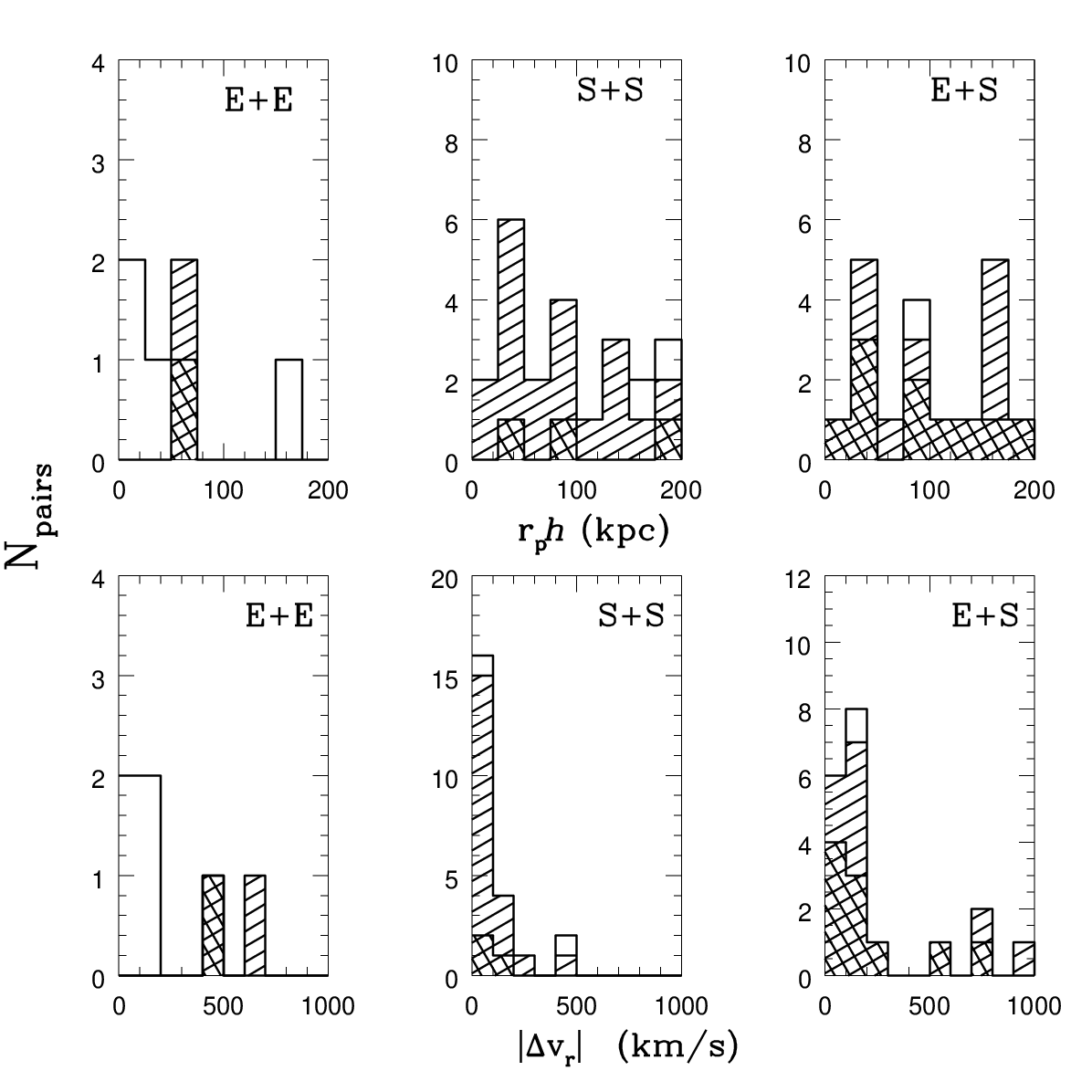}
  \caption{Relation between the presence of
  emission lines in the galaxy spectra and dynamical parameters
  of the pairs. The upper panels show the galaxy-galaxy projected distance  (r$_p$) distribution, 
the lower panels the velocity difference ($|$ $\Delta v_r$ $|$) between galaxies (in each pair) for
  E+E, S+S, E+S respectively. Pairs having at least one galaxy with emission
  spectrum are represented by dashed histogram. The single dash
  indicates pairs in which both members have an emission line spectrum, the double dash pairs
  hosting only one emission line galaxy.  }
    \end{figure}

Figure 2 shows the absolute magnitude (M$_B$ - 5 log $h$)
distribution of ellipticals (T $<$ -3,  upper left), S0s ( -3 $\le $
T $<$ 0,  upper right), early ( 0 $\le $ T $<$ 4, lower left) and
late  ( T $\ge 4$ , lower right) spirals. (M$_B$ has been derived
from B which is available in LEDA, Paturel et al. 2003, for 95/96
galaxies of our sample).
 In analogy with Fig.1
the continuous distribution refers to the whole samples, the hatched
one to galaxies with emission lines. The lower panels of Fig. 2
evidence an extremely large fraction of emission line galaxies
in early and late spirals: 84 \%  (37/44) and 95 \% (19/20), respectively.
 Curiously ellipticals (upper left) also show an
unusually large content (67 \%) of emission line galaxies, but the
statistics are  too poor
and  do not allow us to draw definitive conclusions on this
point. S0 galaxies, instead, are quite abundant in our sample
(22/96) and appear to host a much more limited fraction (23 \%)
of emission line galaxies. From Fig. 2 we see that our sample is
numerically dominated by  early spirals (47 \%), has an almost equal
content of late spirals  (21 \%) and S0s (23 \%) and contains a
limited fraction (9 \%) of ellipticals. In terms of luminosity
(M$_B$) S0s are characterized by a larger content of "faint"
galaxies, about half of them (12/22, 55 \%)  have  M$_B$ $\ge$ -19.5
+ 5 log $h$ %
There are no ellipticals with
M$_B$ larger than that value, while early and late spirals
(with  M$_B$ $\ge$ -19.5
+ 5 log $h$) represent,
respectively, 25 \% and 15 \% of each whole population.
 Further
investigation is required to establish whether the population of low
luminosity S0s is typical of bright galaxy pairs and if the lack of
emission features in these galaxies relates to their low luminosity
either.   

Figure 3 shows the distribution of galaxy-galaxy projected distance
(r$_p$, upper panels) and velocity separation  ($|$ $\Delta v_r$$|$, lower
panels) for E+E , S+S  and E+S  pairs. The continuous distribution
indicate the whole samples, the dashed distributions indicate pairs
having at least one member with emission line spectrum (double dash)
and pairs in which both members have emission lines (single dash).
The fraction of pairs having both members "active" is clearly larger
in S+S (78 \%, 18/23) than in E+S  (42 \%, 8/19). This difference is
induced by the early-type galaxy content of the E+S pairs . The
fractions become, in fact, similar (91 \% and 95 \%) when considering
S+S and E+S pairs in which at least one member has an emission line
spectrum.

Previous work (Barton et al.2000; Alonso et al. 2004) has claimed an increase
of emission line galaxies in pairs with decreasing member projected distances.
In our sample the fraction of emission line galaxies in E+S and S+S pairs
is so high,
 at all member distances, that we hardly  see such an effect.
We stress however that UZC-BGP is a volume limited sample of  bright
isolated galaxy pairs quite different from magnitude limited samples
of close pairs belonging to any kind of environment as are the
samples on which Barton et al. (2000) and Alonso et al. (2004)
evidenced the distance effect. The very large fraction of emission
line galaxies in our sample at all pair distances  may actually
indicate that in isolated pairs of luminous (M$_{Zw}$ $\le$ -18.9 +
5 log $h$) galaxies interaction is at work and effective up to 200
$h$ $^{-1}$ kpc. Both Barton et al. (2000) and Alonso et al.(2004)
find the  emission line enhancement on a much smaller scale ($\sim$
30 $h$ $^{-1}$ kpc). Their result is likely to indicate that when
galaxy pairs are surrounded by companions of comparable luminosity
galaxy-galaxy interaction becomes effective only at very small
distances.\,Our result indicates that fainter companions which
may be present in some UZC-BGP systems in the close (200 $h$
$^{-1}$)  kpc and/or large (1 $h$ $^{-1}$) Mpc surrounding area do
not play an action comparable to the one of luminous member
galaxies.

 The lower panels of Fig.3 indicate a "dynamical" difference  between pairs
  containing  only spirals (middle panel) and pairs containing either both
  (left panel) or at least one  (right panel) early-type  galaxy, the
 former being characterized by a narrower $|$ $\Delta v_r$ $|$ distribution than the latter ones
 (KS confidence level 93.9 \%  and 99.9 \% respectively) .
 This difference (already outlined in paper I) suggests that E+S and E+E pairs might actually
  be embedded within large loose structures as
  their broader $|$ $\Delta v_r$ $|$  distribution  would indicate the presence of a larger potential
  well. In this framework  the emission spectrum of the 3
  early-type galaxies belonging to the E+E pairs could arise from infalling of these
  galaxies within a loose group as suggested from their large $|$ $\Delta v_r$ $|$  value. We have checked
  this hypothesis and found that neither UZC-BGP 51 nor UZC-BGP 77 are part of any
  known galaxy group. However, inspection of available (LEDA) redshifts of galaxies in the surrounding
  environment of both pairs show that UZC-BGP 51 might be part of a loose galaxy group  which
  could possibly be infalling on the nearby (4 $h$ $^{-1}$ Mpc ) Coma cluster; while  UZC-BGP 77A
  (NGC 6018) might be infalling on a galaxy loose group having the same radial velocity of UZC-BGP 77B
  (NGC 6021). More investigation is required to confirm our hypothesis.

\section{Nuclear activity classification}

 \begin{table*}
\begin{center}
\caption[] {Spectral line ratios for standard diagnostic diagrams (diags. 1, 2 and 3).}
\begin{tabular}{||r|c|l|l|l|l||}
\hline
\hline
UZC-BGP & Pair type & [OIII]$_{5007}$/H$_{\beta}$ & [NII]$_{6583}$/H$_{\alpha}$
&[SII]$_{(6717+6731)}$/H$_{\alpha}$ & [OI]$_{6300}$/H$_{\alpha}$\\
\hline
\hline
2A & S+S & 0.49 $\pm$ 0.06 & 0.553 $\pm$ 0.006 & 0.333 $\pm$ 0.004 & 0.043 $\pm$ 0.002 \\
9A & S+S & 1.59 $\pm$ 0.06 & 0.198 $\pm$ 0.003 & 0.382 $\pm$ 0.006 & 0.035 $\pm$0.006\\
14A & E+$\underline{S}$ & 0.516 $\pm$ 0.022&  0.349 $\pm$ 0.006 & 0.278 $\pm$ 0.023 & 0.022 $\pm$ 0.008 \\
20A & E+$\underline{S}$ & 1.32 $\pm$ 0.06 & 0.30 $\pm$  0.01& 0.36 $\pm$  0.02 &  0.023 $\pm$ 0.002\\
24A & S+S & 1.61 $\pm$  0.09&  0.55 $\pm$  0.02 & 0.25 $\pm$  0.03 &  0.045
$\pm$  0.000\\
28A & S+S &2.3 $\pm$ 0.4  & 1.16 $\pm$  0.02 & 0.70 $\pm$  0.05 & 0.40 $\pm$  0.04\\
28B & S+S & 0.67 $\pm$ 0.02& 0.276 $\pm$ 0.002 & 0.19 $\pm$  0.01 & 0.023 $\pm$  0.006  \\
29B & S+S & 0.60 $\pm$ 0.05 & 0.287 $\pm$  0.004 & 0.20 $\pm$ 0.01 & 0.021 $\pm$ 0.003\\
68A & S+S &  0.79 $\pm$  0.02 & 0.25 $\pm$  0.01 & 0.21 $\pm$ 0.02 & 0.044 $\pm$  0.007  \\
74A & S+S & 0.92 $\pm$  0.03 & 0.35  $\pm$  0.02  & 0.22  $\pm$ 0.01 & 0.54  $\pm$  0.03 \\
74B & S+S &1.6$\pm$ 0.1 &  0.18 $\pm$0.02 &  0.232 $\pm$0.003 & 0.05 $\pm$0.01 \\
81B & S+S & 0.36 $\pm$ 0.04& 0.509 $\pm$0.009 & 0.17 $\pm$0.02  & 0.028 $\pm$0.003 \\
83A & S+S & 4.4 $\pm$0.3 & 0.332 $\pm$0.003&0.144 $\pm$0.006  &  0.078$\pm$ 0.005\\
83B & S+S & 0.89 $\pm$0.03&0.262 $\pm$0.006 & 0.26 $\pm$0.01 & 0.044 $\pm$0.003 \\
5A &E+$\underline{S}$ & 1.45 $\pm$ 0.02 & 0.387 $\pm$ 0.005  & 0.248 $\pm$ 0.005 & -\\
22A & S+S & 0.79 $\pm$ 0.02& 0.38 $\pm$ 0.01& 0.17$\pm$ 0.01& -\\
22B  & S+S & 0.64 $\pm$ 0.03&0.33 $\pm$ 0.01 & 0.28 $\pm$ 0.02 & -\\
29A & S+S & 0.28 $\pm$ 0.04& 0.442 $\pm$ 0.004  & 0.102 $\pm$ 0.008 & -\\
66A & S+S & 2.3 $\pm$ 0.4 &0.27 $\pm$ 0.01 &0.198 $\pm$ 0.004 & -\\
69B &S+S & 0.79 $\pm$ 0.02 & 0.76$\pm$ 0.02 & 0.51 $\pm$ 0.03 & - \\
81A& S+S& 3.0 $\pm$ 0.3  & 0.203 $\pm$ 0.004 & 0.18 $\pm$ 0.02 & -\\
82A& S+S & 0.42 $\pm$ 0.04 & 0.386 $\pm$ 0.004  & 0.18 $\pm$ 0.02 & -\\
44A & S+S & 1.3 $\pm$ 0.2 & 0.39 $\pm$ 0.01 & - & -\\
65A &E+$\underline{S}$ & 0.97 $\pm$ 0.05 & 0.263 $\pm$ 0.008   & - & -\\
\hline
\end{tabular}
\end{center}
\end{table*}

To classify nuclear activity in our galaxy sample, we have made use of the
standard diagnostic diagrams (Baldwin et. al. 1981; Veilleux \& Osterbrock 1987; Veilleux 2002),
also known as the BPT diagrams,
which have proved to be an extremely
 efficient method to distinguish the different types of activity encountered
in emission line galaxies (Veilleux et al. 1995; Veron et al. 1997; Goncalves et al. 1999).
Moreover, being based on ratios of emission lines which are very close in wavelength they
are almost unaffected by reddening corrections (Veilleux \& Osterbrock 1987).
The diagnostic diagrams  relate [OIII]$_{5007}$/H$_{\beta}$  to
[NII]$_{6583}$/H$_{\alpha}$ , [SII]$_{(6717+6731)}$/H$_{\alpha}$
and [OI]$_{6300}$/H$_{\alpha}$.

Of the 68 galaxies with emission spectra (cfr. Tables 1,2 and 3), 24
allowed us to build up from one to three of the above mentioned
standard diagnostic diagrams. These galaxies are listed in Table 4,
in order of decreasing number of diagnostic diagrams, i.e. the first
14 galaxies have all line ratios measured, the subsequent 8 galaxies
only three line ratios and the remaining 2 only two. Table 4,
reports, for these galaxies, UZC-BGP identificator (column 1), pair/galaxy 
morphology (column 2,  galaxy morphology is underlined in cases of 
mixed pair morphology), line ratios and errors (column 3,4,5,6). Each
line ratio has been obtained averaging line ratios obtained
independently by each of us and the associated error represent the
standard deviation ($\sigma$).
 The vast majority
(20/24) of  galaxies listed in Table 4 belongs to the S+S pair sample
 and the remaining 4 are however spirals, implying that there are
no early-type galaxies in our sample showing at least 4 emission lines.

 \begin{figure}
   \centering
  \includegraphics[width=8cm]{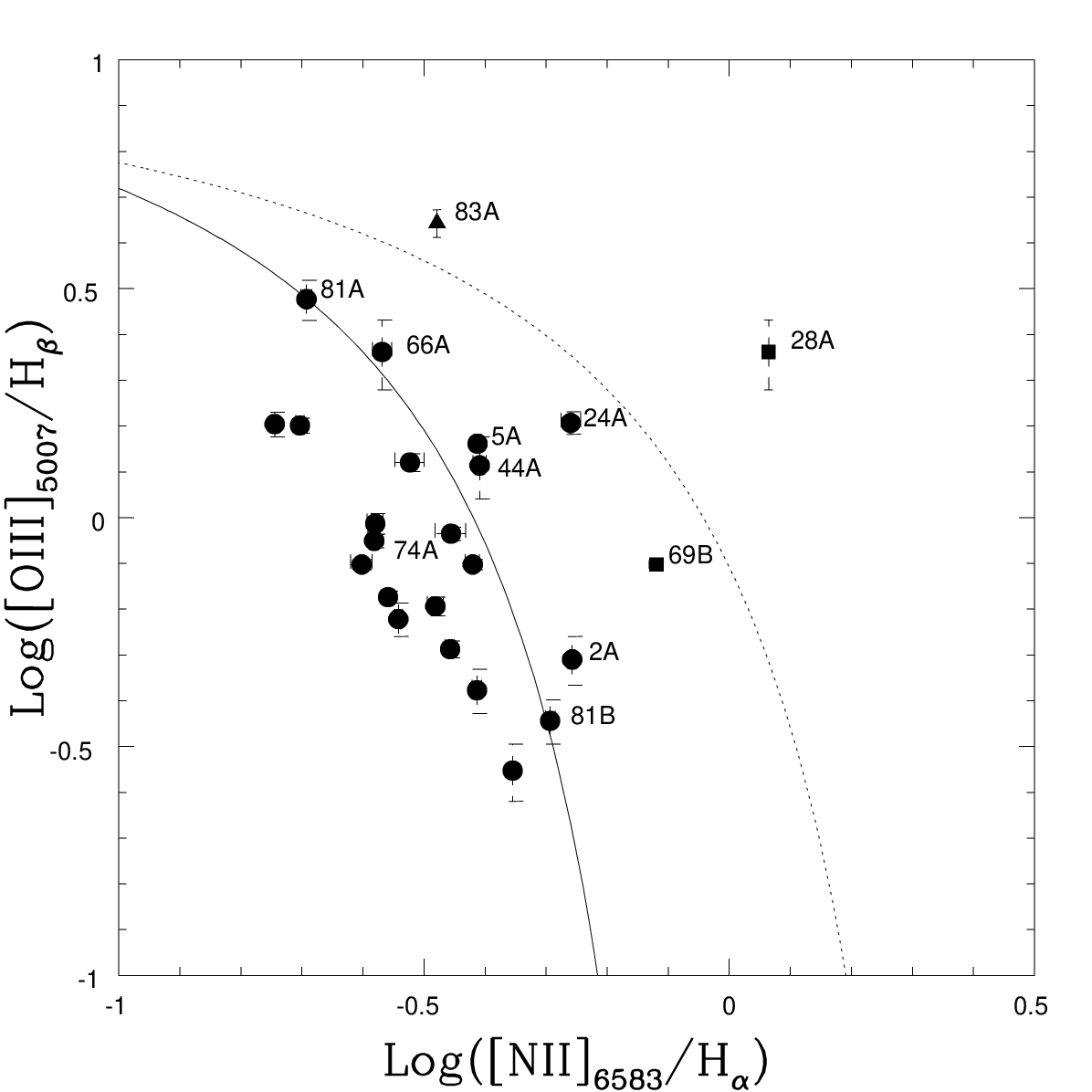}
  \caption{Log ([OIII]$_{5007}$/H$_{\beta}$)  versus Log ([NII]$_{6583}$/H$_{\alpha}$)
  diagnostic diagram (diag. 1). Circles, squares and triangle indicate
  classification into SB, LINER and Sy 2 according to VO87 scheme. The
  solid curve on the left is the Kauff03 sequence,while
  the dotted curve on the right
  corresponds to
  the Kew01 sequence. Both sequences are supposed to
  separate SB from AGNs. Identification is provided for all galaxies
  above Kauff03 sequence as, according to it, they should all be
  classified as AGNs. One galaxy (UZC-BGP 74A) below the Kauff03 sequence has
   been identified too
  as it  lays in the LINER region  of diag. 3.
  }
    \end{figure}

The Log([OIII]$_{5007}$/H$_{\beta}$)  versus Log
([NII]$_{6583}$/H$_{\alpha}$) diagram (hereafter diag. 1), for all
the galaxies listed in Table 4, is shown in Fig. 4. Different
symbols indicate different nuclear activity according to Veilleux \&
Osterbrock (1987) empirical classification (hereafter VO87)
 which separates high  from low excitation galaxies (Seyfert 2 from LINERs
and HII galaxies from SBs) based on a [OIII]$_{5007}$/H$_{\beta}$ value of 3 and LINER from SBs based on a
 [NII]$_{6583}$/H$_{\alpha}$ value of 0.6.

    \begin{figure}
   \centering
  \includegraphics[width=8cm]{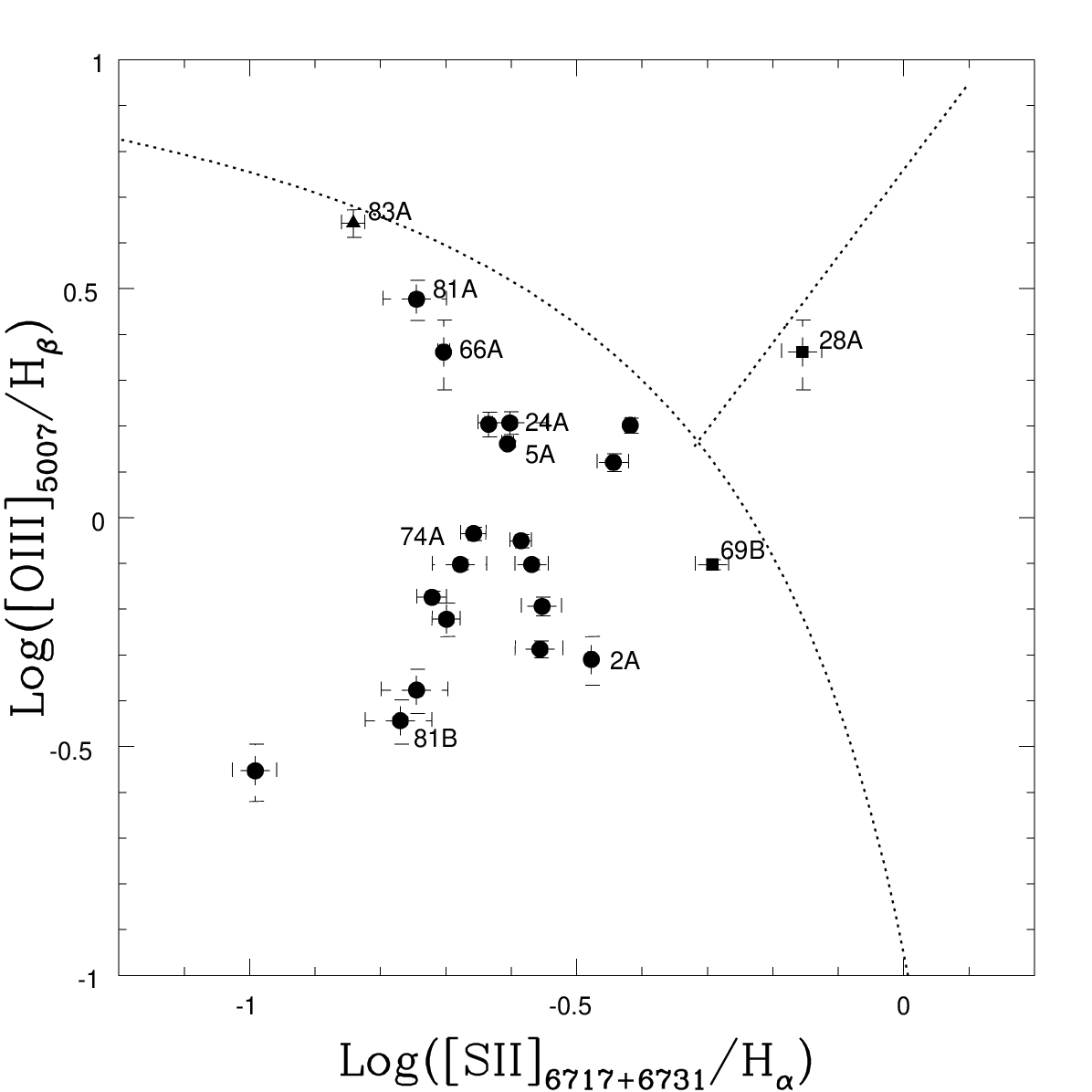}
  \caption{Log ([OIII]$_{5007}$/H$_{\beta}$)  versus Log ([SII]$_{(6717+6731)}$/H$_{\alpha}$)
  diagnostic diagram (diag. 2). In analogy with Fig. 4, circles, squares
   and triangle indicate classification into SB, LINER and Sy 2 according
   to VO87 scheme.
  The dotted curve represent
  the Kew01 sequence separating SB (below)  from AGN (above) region. The dotted line
  is the Kew06 sequence which separate Seyfert (above) from LINERs
  (below).
  Identified on this plot are all the
  galaxies above the Kauff03 sequence in diag.1 and UZC-BGP 74A.
    }
    \end{figure}

According to VO87, our sample contains one Seyfert 2 galaxy
 (triangle), two LINERs (squares) and 21 SBs (circles). The Sy 2
 (UZC-BGP 83A) and the LINERs (UZC-BGP 28A and 69B) are identified in Fig. 4.
 On the same diagram we show two curves, the solid one to the left
 is the Kauffman et al.(2003) sequence (hereafter Kauff03), the dotted
 one to the right is the Kewley et al.
  (2001) sequence (hereafter Kew01). Both sequences are supposed to
  separate SB from AGNs. Kauff03
  is an empirical sequence which has been derived from
  a huge sample ($\sim$ 22 000) of SDSS
  emission line galaxies,
  while Kew01
 is a theoretical sequence derived using a wide set of models
 accounting for photoionization and
 stellar population synthesis.
 Figure 4 shows that there are 8 galaxies
falling between Kauff03 and Kew01
 sequences.
  These galaxies, ordered by decreasing value of  [OIII]$_{5007}$/H$_{\beta}$, are
  UZC-BGP 81A, 66A, 24A, 5A, 44A, 69B (a LINER according to VO87), 2A and 81B and have all been
  identified in Fig. 4. Two of these galaxies (UZC-BGP 81A and 66A) could fall just below
the Kauff03 sequence taking into account the error associated to  the  [OIII]$_{5007}$/H$_{\beta}$ measure,
while another one (UZC-BGP 81B) would move
 from just above the sequence to the sequence itself.
  One further galaxy (UZC-BGP 74A)  is identified in Fig. 4 as
  although it lays, in this diagram, well below the Kauff03 sequence, it
  occupies  the LINER region
  of diag. 3 (Fig. 6)

Figure 5 shows the Log ([OIII]$_{5007}$/H$_{\beta}$)   versus Log([SII]$_{(6717+6731)}$
/H$_{\alpha}$) diagram
(hereafter diag. 2) for the 22
galaxies of Table 4 which have also these line ratios measured. In analogy with Fig. 4 different symbols stand
for different classification in the VO87 scheme ([OIII]$_{5007}$/H$_{\beta}$ $\ge 3$ and
[SII]$_{(6717+6731)}$ /H$_{\alpha}$ $\ge 0.4$  separate,
respectively, Sy 2 and LINER from SBs). VO87 criteria confirm the classification in Sy 2
and LINERs obtained from diag. 1  (Fig. 4) for UZC 83A, 28A and 69B.
 The dotted curve and line correspond to Kew01 and Kew06 (Kewley et al.2006)
 sequences. The former one separates SB (below) from AGNs (above), the latter
 one Sy 2 (above) from LINERs (below). Identified, on this plot, are also
  the three
 confirmed (VO87) AGNs  and 7 (of the 8) galaxies laying between Kauff03 and Kew01
 sequences in diag. 1 (Fig.4), as UZC-BGP 44A has only diag. 1 available (cfr. Table
 4). Finally the position of UZC-BGP 74A is indicated, since
  this galaxy occupies the LINER region of diag. 3 (Fig. 6).

 Figure 5 shows that, according to Kew01 and Kew06 classification, only one
 galaxy (UZC-BGP 28A) would be classified as  LINERs. All the others would be
 SB, although UZC-BGP 83A (Sy 2 following VO87 scheme) lies just below the
 Kew01 sequence and would move exactly on it if one lets
[OIII]$_{5007}$/H$_{\beta}$ attain its maximum possible value (within the error).

    \begin{figure}
   \centering
  \includegraphics[width=8cm]{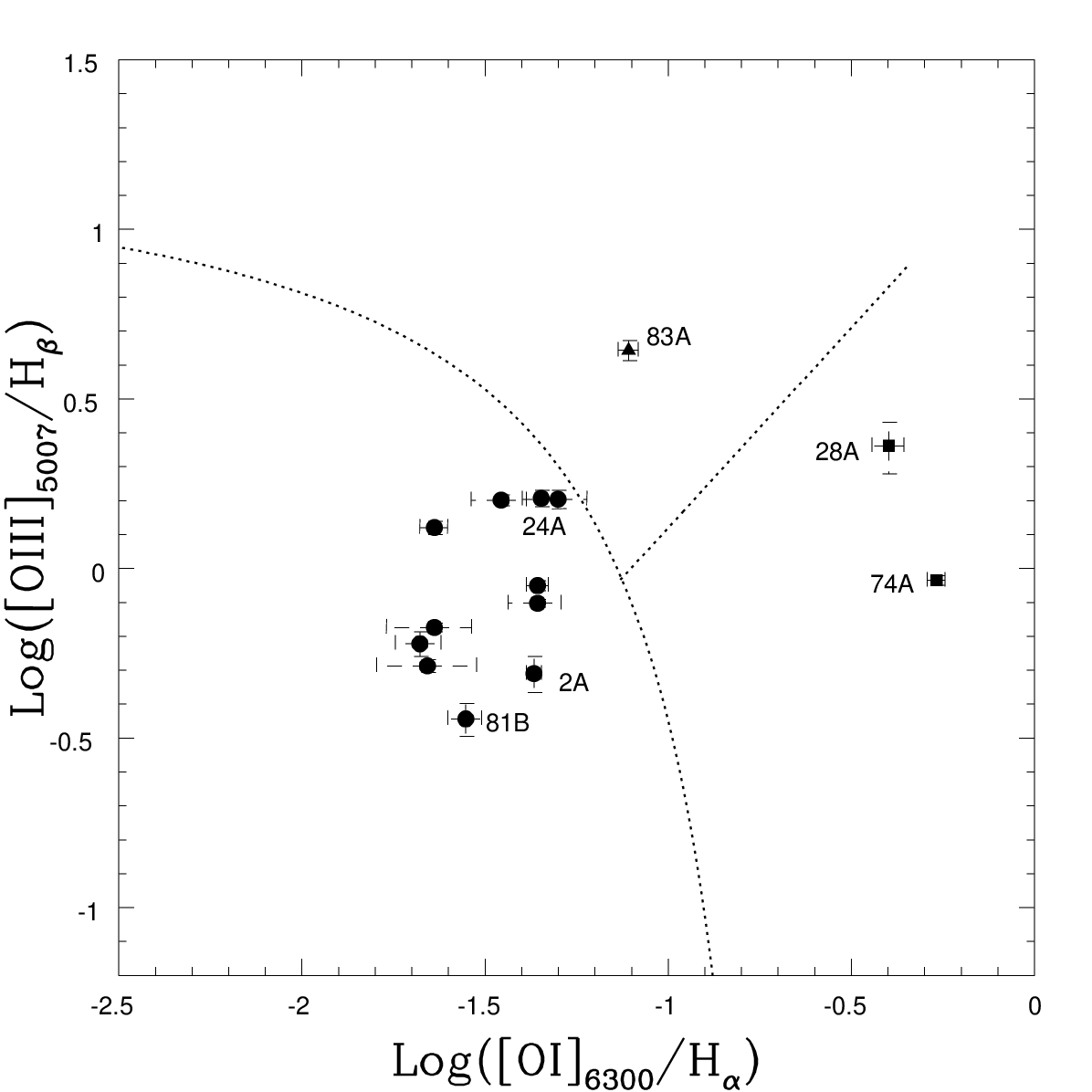}
  \caption{Log ([OIII]$_{5007}$/H$_{\beta}$)  versus Log ([OI]$_{6300}$/H$_{\alpha}$)
  diagnostic diagram (diag. 3). Symbols are as in Fig. 4 and 5.
  In analogy with Fig. 5 the dotted curve and line represent
  Kew01 and Kew06 sequences separating, respectively, SB from AGNs and
  Seyfert from LINERs. Identification is provided
  for all the galaxies lying in the ``AGN'' region either in this or in another diagnostic
diagram.}
    \end{figure}


Figure 6 shows the Log ([OIII]$_{5007}$/H$_{\beta}$) versus
Log ([OI]$_{6300}$/H$_{\alpha}$) diagram (hereafter diag. 3) for the
14 galaxies in Table 4 having these line ratios measured.
In analogy with Figs. 4 and 5 different symbols represent different nuclear activity types according to VO87
scheme which separates Sy 2 from SB and LINERs based on a value of  [OIII]$_{5007}$/H$\beta$ of 3 and
requires a value of
[OI]$_{6300}$/H$\alpha$ $\ge 0.08$ for Sy 2  and $\ge 0.17$ for LINERs.
Here too the dotted curve and line correspond to Kew01 and Kew06 sequences
separating SB from AGNs and Sy 2 from LINERs respectively.
 Identified, in this plot, are all the galaxies lying above Kew01 curve and
 3 (over 8) galaxies lying in the Kauff03 Kew01 region of diag. 1 (Fig. 4).
 Two galaxies (UZC-BGP 28A and 74A) occupy the Kew01 -- Kew06 LINER region, 
 one (UZC-BGP 83A) the Sy 2 region. This classification is confirmed by VO87.
 One galaxy  in Fig. 6 (UZC-BGP 74B) could move from the SB locus to the Kew01 SB/AGN separation
curve when taking into account the error associated to the [OI]$_{6300}$/H$_{\alpha}$ measure. This galaxy is  very close to UZC-BGP 24A (at its right),
 but we have not identified it in Fig. 6, as it occupies the SB region both in diags. 1 and 2.

\begin{table*}
\begin{center}
\caption[] {Spectral line ratios for diag. 4 }
\begin{tabular}{||r|c|l|l||}
\hline
\hline
UZC-BGP & Pair type &[NII]$_{6583}$/H$_{\alpha}$ &[NII]$_{6583}$\\
\hline
\hline
3A & E+$\underline{S}$ & 0.45 $\pm$ 0.01& 5.1  $\pm$ 0.8\\
4B  & E+$\underline{S}$ & 2.99 $\pm$ 0.01 & 1.01 $\pm$ 0.01 \\
8B & S+S & 0.34 $\pm$ 0.02 & 4.2  $\pm$ 0.3 \\
9B  & S+S & 0.332 $\pm$ 0.004& 3.71 $\pm$ 0.06\\
10A & $\underline{E}$+S &1.27 $\pm$ 0.08 &  2.11 $\pm$ 0.04\\
10B  & E+$\underline{S}$ & 0.85 $\pm$ 0.02 &  2.9 $\pm$ 0.1\\
17B  & S+S & 0.337  $\pm$ 0.004& 5.97 $\pm$  0.04\\
18A  & E+$\underline{S}$ & 0.245 $\pm$ 0.007& 12.0 $\pm$  0.4 \\
20B  & $\underline{E}$+S & 2.9 $\pm$0.1 & 2.9 $\pm$ 0.1\\
21B  & E+$\underline{S}$ & 9.4 $\pm$ 1.6&  2.8 $\pm$ 0.2\\
24B  & S+S & 0.61 $\pm$ 0.02 &  6.4 $\pm$ 0.3\\
26B  & $\underline{E}$+S & 2.8 $\pm$ 0.8 & 1.3 $\pm$ 0.3\\
27A  & $\underline{E}$+S & 3.0 $\pm$  0.4 &  4.76 $\pm$  0.07\\
27B  & E+$\underline{S}$ & 0.50 $\pm$ 0.04 & 4.4 $\pm$ 0.2\\
36A  & S+S & 2.5 $\pm$ 0.4& 3.5 $\pm$ 0.6 \\
36B & S+S & 0.58 $\pm$ 0.02 & 5.8 $\pm$ 0.2\\
37A  & S+S & 0.70 $\pm$ 0.03  & 3.6 $\pm$ 0.3\\
37B  & S+S & 0.36 $\pm$ 0.02 & 5.5 $\pm$ 0.5\\
39A &  $\underline{E}$+S & 0.838 $\pm$ 0.006& 6.4 $\pm$ 0.1\\
44B  & S+S & 0.57 $\pm$ 0.02 & 6.2 $\pm$ 0.2\\
49A  & E+$\underline{S}$ & 0.375 $\pm$ 0.005 & 7.7 $\pm$ 0.3\\
59B  & S+S & 0.650 $\pm$ 0.008 & 6.10 $\pm$ 0.07\\
61A  & E+$\underline{S}$ & 0.752 $\pm$ 0.07   & 7.8 $\pm$ 0.3 \\
68B  & S+S & 0.36 $\pm$ 0.01 & 4.0 $\pm$ 0.3\\
71B  & S+S & 3.2 $\pm$ 0.1& 3.8 $\pm$ 0.2\\
75A  & $\underline{E}$+S & 1.565 $\pm$ 0.007 & 2.55 $\pm$ 0.07\\
75B  & E+$\underline{S}$ & 0.95 $\pm$ 0.01 & 2.7 $\pm$ 0.3 \\
77A  & E+E & 0.74 $\pm$ 0.03 & 4.5 $\pm$ 0.3\\
79A  &  $\underline{E}$+S & 1.65 $\pm$ 0.06 & 4.4 $\pm$ 0.4\\
80B  & S+S & 0.28 $\pm$ 0.06 & 8.1 $\pm$ 0.3 \\
84B  & E+$\underline{S}$ & 0.44  $\pm$ 0.02  & 4.42  $\pm$ 0.02 \\
\hline
\end{tabular}
\end{center}
\end{table*}

The spectral analysis of the 24 emission line galaxies listed in Table 4
 allow us to classify unambiguously 14 of them as for these galaxies
 all the different classification schemes (V087, Kauff03 and  Kew01-06) 
give consistent results on the 3 diagnostic
 diagrams. Of the remaining 10 galaxies 8 lay above the Kauff03 sequence and
 below the Kew01 in diag. 1. One (UZC-BGP 44A), of these 8,  has only diag. 1 available,
 4 have also diag. 2 and 3 have both diags. 2 and 3 available. VO87 and Kew01 classification
 schemes agree on SB activity in the other diagnostic diagrams for 6 of the 7 galaxies having
 either both diag. 2 and diag. 3 or only diag. 2 available, only UZC-BGP 69B is classified
 (in diag. 2) LINER in the VO87 scheme and SB in the Kew01 one.
 The 2 other "difficult" cases are  UZC-BGP 83A  and 74A. The first one is a Sy 2 in
 both diags. 1 and 3, for all classification schemes, but in diag. 2  is a SB,
according to Kew01 (although it lays very
close to the SB/AGN border and would sit exactly on  it if [OIII]$_{5007}$/H$\beta$ attains it
maximum permitted value within the error), and a Sy 2  according to VO87. More ``difficult'' is the case of
 UZC-BGP 74A which is
a LINER  in diag. 3 (for both Kew01 and VO87) and a SB in diags. 1 and 2
  for all classification schemes.

Galaxies displaying different kind of nuclear activity, in different diagnostic diagrams,
 should be classified as composite objects (Kewley et al. 2006).
 However, in order to avoid double classification
 (e.g. SB/AGN, Sy2/SB, LINER/SB) we have decided to
 assign to each galaxy  the "most frequent" classification according
 to different schemes applied to the 3 diagnostic diagrams. We keep,
 however, record of each classification
 scheme in each diagnostic diagram in Table 6, which summarizes
 the results of our spectral analysis.

\begin{figure}
   \centering
  \includegraphics[width=8cm]{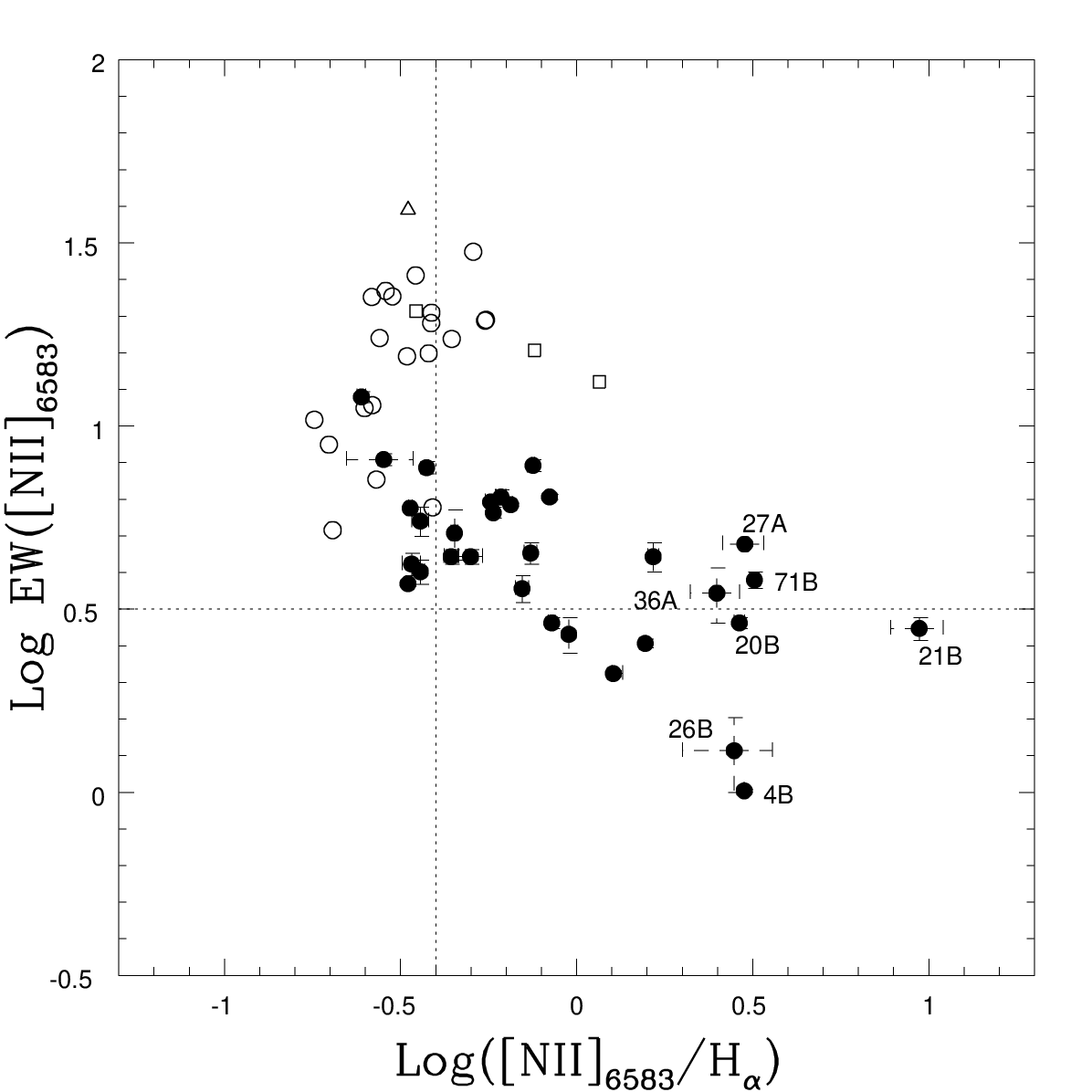}
  \caption{Log EW ([NII]$_{6583}$)  versus Log ([NII]$_{6583}$/H$_{\alpha}$)
  diagnostic diagram (diag. 4)  that we have used to classify
the  31 galaxies listed in Table 5  (filled circles).
   The vertical dotted line separates  the SB locus from the AGN one,
   the horizontal dotted line separates AGNs (up) from LLAGNs (low).
 The open symbols indicate the position, 
in this diagram, of the 24 galaxies listed in Table 4, whose nuclear activity has been classified by means of one
or more standard diagnostic diagrams. (Circles, squares  and triangle indicate classification into SB, LINER 
and Sy 2 according to V087 scheme applied to diags. 1, 2 and 3).
  }
    \end{figure}

A large fraction of emission line galaxies (31/68)  have well defined
H$_{\alpha}$ and [NII] lines but  do  not show [OIII] and/or
H$_{\beta}$ features and thus does not allow us to use the standard
diagnostic diagrams. For these galaxies, listed in Table 5, the
activity type can be classified (Coziol et al. 1998)  comparing the
EW of [NII]$_{6583}$ feature with  the ratio of [NII]$_{6583}$ to
H$_{\alpha}$. These values are listed in Table 5, together with
their errors, which, in analogy with Table 4, represent respectively
the average of 3 measures obtained independently by each of us and
the associated $\sigma$. The underlined morphology in column 2 indicates, as in
Table 4, the ``rough'' galaxy morphology when galaxy pair is of mixed (E+S) type.

Following Coziol et al. (1998) we have classified the  31 galaxies listed in Table 5
as SB if they have  Log ([NII]$_{6583}$/H$_{\alpha}$) $<$ -0.4 and Log EW([NII]$_{6583}$) $\ge 0.5$;
AGN and LLAGN if they have   Log EW([NII]$_{6583}$/H$_{\alpha}$) $\ge - 0.4$
and  Log EW([NII]$_{6583}$) $\ge 0.5$ or $<$ 0.5 respectively.
The lines which separate the loci of SB, AGN and LLAGN
are represented in Fig. 7 which shows the  Log EW([NII]$_{6583}$)  versus Log ([NII]$_{6583}$/H$_{\alpha}$)
 diagram (hereafter diag. 4)
for  the 31 galaxies listed in Table 5 (filled circles).
 Among these galaxies
we have identified the  ones
 (7), showing the largest value of [NII]$_{6583}$/H$_{\alpha}$. This
behaviour is due to the presence of a strong H$_{\alpha}$ feature
in absorption depressing the emission one and resulting in an
enhanced [NII]$_{6583}$/H$_{\alpha}$ ratio. From Fig. 7 we see that only one galaxy (UZC-BGP 36A) could have its classification
changed (from AGN to LLAGN) when taking into account the error associated to  EW([NII]$_{6583}${\bf )} measure.

 The classification scheme adopted by Coziol et al. (1998) is empirical and might, thus,  be questioned, 
for this reason we also show, on the same plot (Fig. 7), the 24 galaxies (open symbols) listed in Table 4, that we have
classified by means of one or more standard diagnostic diagrams. The different symbols 
indicate (in analogy to Figs. 4, 5, and 6) different nuclear activity kind.
 From Fig. 7 we see that  for low values of Log EW([NII]$_{6583}$) ( $\sim < $ 1.1)  Coziol classification scheme
holds well, since all galaxies classified as SB (open circles) occupy the SB region of the diagram, whereas
for larger values of  Log EW([NII]$_{6583}$) there are SBs in the AGN region and
AGN in the SB ones \footnote{ This behaviour is 
not unexpected since the sequence which separates AGN from SBs, in Fig. 4, is bended in  [NII]$_{6583}$/H$_{\alpha}$ and
Sy 2 are characterized by lower values of [NII]$_{6583}$/H$_{\alpha}$ than LINERs. }. The agreement between Coziol and standard classification, occurring at low values of Log EW([NII]$_{6583}$),
where SBs from Tables 4 and 5 overlap, make us to feel rather confident about this classification scheme. 

Of the remaining 13 galaxies with emission lines (cfr. Tables 1, 2 and
3), 10  show only the [NII]$_{6583}$ emission feature and thus, according
to Coziol et al. (1998) we have classified them as LLAGN candidates,
while 3 show only H$_{\alpha}$  in emission and have not been
classified.

\begin{table*}
\begin{center}
\caption[] {Nuclear Activity classification}
\begin{tabular}{||r|l||l|l|l||l|l||l|l||l|l|l||}
\hline
UZC-BGP & Pair type &\multicolumn{3}{c|}{Diag. 1} &\multicolumn{2}{c|}{Diag. 2}
&\multicolumn{2}{c|} {Diag. 3} & Diag. 4  & Adopted
&Known\\
\cline{3-9}
& & VO87 & Kew01 & Kauff03 &VO87 &Kew01 &VO87 & Kew01 & & &  \\
\hline
\hline
2A & S+S & SB & SB & AGN  & SB & SB & SB & SB &   & SB & SB\\
2B & S+S &-- &-- &-- &-- &--   &--  & -- &   & LL: & \\
3A & E+$\underline{S}$ & --&-- &-- &-- &-- &-- &-- & AGN & AGN & LINER\\
4B  & E+$\underline{S}$ & --& --& --& -- & --& --& --& LLAGN   &  LLAGN &\\
5A &E+$\underline{S}$ & SB & SB & AGN & SB & SB & --& -- &  &  SB & HII\\
8B & S+S & --& --& --&-- & --& --& --& SB  &   SB &\\
9A & S+S & SB & SB & SB  & SB & SB & SB & SB &   &SB & \\
9B & S+S & --& --& --& -- & --& --& -- & SB &  SB &\\
10A& $\underline{E}$+S & --& --& --&-- & --& --& --& LLAGN &  LLAGN &\\
10B& E+$\underline{S}$ & --& --& --&-- & --& --& --& LLAGN   & LLAGN &\\
14A& E+$\underline{S}$&  SB & SB & SB  & SB & SB & SB & SB &    & SB & SB\\
16B & E+$\underline{S}$  & --& --& --&-- & --& --& --&   &  LL: &\\
17B &S+S & -- &--&-- &--&--  &-- & --  & SB    & SB &\\
18A & E+$\underline{S}$ & -- &--&-- &--&--  &-- & --  & SB &  SB &LLIRG\\
20A &E+$\underline{S}$ & SB & SB & SB  & SB & SB & SB & SB &   &SB &  \\
20B &$\underline{E}$+S & -- &--&-- &--&--  &-- & --  & LLAGN   & LLAGN &\\
21B& E+$\underline{S}$ & -- &--&-- &--&--  &-- & --  &LLAGN   & LLAGN &\\
22A & S+S & SB & SB & SB &SB & SB & -- & --  &     & SB &\\
22B& S+S & SB & SB & SB &SB & SB &-- & --&  &  SB &\\
24A & S+S & SB & SB & AGN  & SB & SB & SB & SB &   & SB  & \\
24B & S+S & -- &--&-- &--&--  &-- & --  &AGN  &   AGN  &\\
26A&E+$\underline{S}$ & -- &--&-- &--&--  &-- & --  &  &   LL: &\\
26B& $\underline{E}$+S & -- &--&-- &--&--  &-- & --  & LLAGN    & LLAGN & \\
27A& $\underline{E}$+S & -- &--&-- &--&--  &-- & --  & AGN  &   AGN &\\
27B& E+$\underline{S}$ & -- &--&-- &--&--  &-- & --  & AGN &   AGN &\\
28A & S+S & LINER & LINER & AGN  & LINER & LINER & LINER &LINER & &     LINER & LINER \\
28B & S+S & SB & SB & SB  & SB & SB & SB & SB &  &  SB &\\
29A & S+S & SB & SB & SB  & SB & SB & -- & -- &  &  SB &\\
29B  & S+S & SB & SB & SB  & SB & SB & SB & SB &  &  SB &\\
36A&S+S & -- &--&-- &--&--  &-- & --  &AGN &   AGN &\\
36B &S+S & -- &--&-- &--&--  &-- & --  &AGN  &  AGN &\\
37A &S+S & -- &--&-- &--&--  &-- & --  &AGN  &  AGN & NLAGN\\
37B &S+S & -- &--&-- &--&--  &-- & --  & SB &  SB &\\
39A& $\underline{E}$+S & -- &--&-- &--&--  &-- & --  & AGN &   AGN &\\
39B&E+$\underline{S}$ & -- &--&-- &--&--  &-- & --  &  &  LL: &\\
44A&S+S &SB & SB & AGN &-- & -- & -- & -- &    & SB &\\
44B&S+S& -- &--&-- &--&--  &-- & --  & AGN  &   AGN  &\\
49A &E+$\underline{S}$ & -- &--&-- &--&--  &-- & --  & SB &   SB &\\
51B & E+E& -- &--&-- &--&--  &-- & --  & &   LL: &\\
59A & S+S & -- &--&-- &--&--  &-- & --  &   &
LL: &\\
59B & S+S & -- &--&-- &--&--  &-- & --  & AGN  &  AGN &\\
61A & E+$\underline{S}$ & -- &--&-- &--&--  &-- & --  & AGN  &  AGN &\\
65A & E+$\underline{S}$& SB & SB & SB & -- & -- &-- & --&  & SB &\\
66A & S+S &SB & SB & AGN & SB &SB & -- & --&   & SB &\\
68A & S+S & SB & SB & SB  & SB &SB & SB & SB &    & SB &\\
68B & S+S & -- &--&-- &--&--  &-- & --  &SB &  SB & SB\\
69B & S+S & LINER & SB & AGN  & LINER & SB &-- & -- &   &   LINER & LLIRG\\
71A & S+S & -- &--&-- &--&--  &-- & --  &  &  LL: &\\
71B & S+S & -- &--&-- &--&--  &-- & --  & AGN  & AGN &\\
74A & S+S & SB & SB & SB &SB & SB & LINER & LINER  &   & SB &\\
74B & S+S & SB & SB & SB &SB & SB & SB & SB & & SB &\\
75A & $\underline{E}$+S & -- &--&-- &--&--  &-- & --  & LLAGN &   LLAGN &\\
75B & E+$\underline{S}$ & -- &--&-- &--&--  &-- & --  & LLAGN &   LLAGN &\\
77A & E+E & -- &--&-- &--&--  &-- & --  & AGN &   AGN &\\
77B & E+E & -- &--&-- &--&--  &-- & --  & &  LL: &\\
79A & $\underline{E}$+S & -- &--&-- &--&--  &-- & --  &AGN &   AGN & Sy 2\\
79B & E+$\underline{S}$ & -- &--&-- &--&--  &-- & --  & &   LL: &\\
\hline
\end{tabular}
\end{center}
\end{table*}

\setcounter{table}{5}
\begin{table*}
\begin{center}
\caption[] {Nuclear Activity classification}
\begin{tabular}{||r|l||l|l|l||l|l||l|l||l|l|l||}
\hline
UZC-BGP & Pair type &\multicolumn{3}{c|}{Diag. 1} &\multicolumn{2}{c|}{Diag. 2}
&\multicolumn{2}{c|} {Diag. 3} & Diag. 4  & Adopted
&Known\\
\cline{3-9}
& & VO87 & Kew01 & Kauff03 &VO87 &Kew01 &VO87 & Kew01 & & &  \\
\hline
\hline
80B& S+S  & -- &--&-- &--&--  &--  & --   & SB &   SB &\\
81A & S+S& SB & SB & SB/AGN &SB & SB & -- & -- &  & SB & \\
81B& S+S& SB & SB & AGN  &SB & SB & SB & SB &  & SB &  \\
82A & S+S& SB & SB & SB  &SB & SB  &--  & --&  &  SB &\\
83A &S+S& Sy 2 & AGN & AGN & Sy 2 & SB & Sy 2 & Sy 2&  & Sy 2   & Sy 2\\
83B &S+S& SB & SB & SB &SB & SB & SB & SB &  &  SB &\\
84A & $\underline{E}$+S  & -- &--&-- &--&--  &-- & --  & &   LL: &\\
84B & E+$\underline{S}$  & -- &--&-- &--&--  &-- & --   &AGN &    AGN & \\

\hline
\end{tabular}
\end{center}
\end{table*}

Table 6 summarizes the nuclear activity classification for 65/68
emission line galaxies of our sample. The classification has been
performed on the basis of one, or more, diagnostic diagrams for 55
galaxies and on the presence of the unique emission feature
[NII]$_{6583}$ , for 10 , that following Coziol et al.
(1998) we have classified as  LLAGN candidates. Three galaxies
(UZC-BGP 17A , 80A and 82B) show only H$_{\alpha}$ in emission and thus
could not be classified and included in Table 6. Besides UZC-BGP
identificator (column 1) and pair/galaxy morphology (column 2, as in
Tables 4 and 5 galaxy morphology is underlined in  E+S pairs), Table 6
lists the classification derived from diags. 1, 2 and 3
(columns 3, 4 and 5), according to VO87 (first subcolumn), Kew01
(second subcolumn) and Kauff03 (third subcolumn, only for diag. 1).
In column 6 the classification into AGN, SB or LLAGN, according to
diag. 4, is given. Column 7 reports the classification that we have
adopted for each galaxy ( LL: stands for LLAGN candidate, based on
the unique presence of [NII]$_{6583}$  feature).
 Finally,
column 8 reports the activity type classification available, for 11 galaxies, from NED.

Comparison of column 7 with column 8 shows a very good agreement between our classification
 and the available one. The agreement is very good
 also for the cases (UZC-BGP 2A, 5A and 83A) in which we have adopted the "majority"  classification
 criterion and when the classification has been based only on   H$_{\alpha}$ and [NII]$_{6583}$
  (UZC-BGP  18A and 37A).  Only for UZC-BGP 79A (NGC 6251) we could not confirm the Sy 2
  nature as
 we did not detect  H$_{\beta}$ \footnote {An extremely  faint H$_{\beta}$ emission, largely affected
 by an underlying strong absorption, is
actually present in in the spectrum of this galaxy, giving a
[OIII]$_{5007}$/H$_{\beta}$  value of 3.20 ($\pm 0.40$). This last
one, coupled to the  [NII]$_{6583}$/H$_{\alpha}$ value of Table 5,
would locate UZC-BGP 79A in the (VO87) Sy 2 region of diag. 1.
However, since  H$_{\beta}$ feature is largely below our adopted
threshold we maintain for this galaxy the AGN classification that we
have derived on the basis of the  [NII]$_{6583}$ and H$_{\alpha}$
features only (diag. 4).}(cfr. Table 3)  at our imposed
threshold limit for emission lines (S/N $\ge 5$).

From Table 6 we see that among the 65  galaxies with classified nuclear activity,\, 29 are SB,
18  AGN ( including 2 LINERs and 1 Sy 2), 8 LLAGN and   10  LLAGN candidates.
SB is the most common kind of nuclear activity
encountered in our sample (30 \% of galaxies display it),
while AGN  is limited to a smaller fraction
(19 \%) of galaxies. SBs are more commonly found in S+S pairs  than
AGNs. There are, in fact,  23/29 SB galaxies and 10/18 AGNs in S+S
pairs (these last 10 include the 2 LINERs and the Sy 2), which implies a fraction
 of SB and AGN per spiral galaxy  in S+S pairs
of  50 \% and  22 \% respectively. The remaining 6 SBs are all
hosted in the spiral member of E+S pairs, while the 8 AGNs equally
divide between spirals and early-type galaxies, one (UZC-BGP 77A) of
these last being actually hosted in a E+E pair. As a whole the
fractions of SB and AGNs per spiral galaxy are  45 \% and 22 \%
respectively.  LLAGNs, instead, are exclusively found in E+S pairs
and in half (4/8) cases this kind of activity is displayed by the
early-type galaxy  member of the pair. The distribution of LLAGN
candidates (LL: in column 7 of Table 6) appears somewhat different
from the LLAGN one, there are, in fact,  3/10  LL: in S+S pairs
and of the remaining 7 LL: only 3 are early-type galaxies. It
is difficult to state whether the difference between LLAGNs and
LLAGN candidates confirms previous finding (Ho et al. 1994)
concerning the rather heterogenous nature of LLAGNs, as our
classification in LLAGN candidates has been based on the presence of
the [NII]$_{6583}$ feature alone. 


\section{The relation between nuclear activity and host galaxy}

\begin{figure}
   \centering
  \includegraphics[width=8cm]{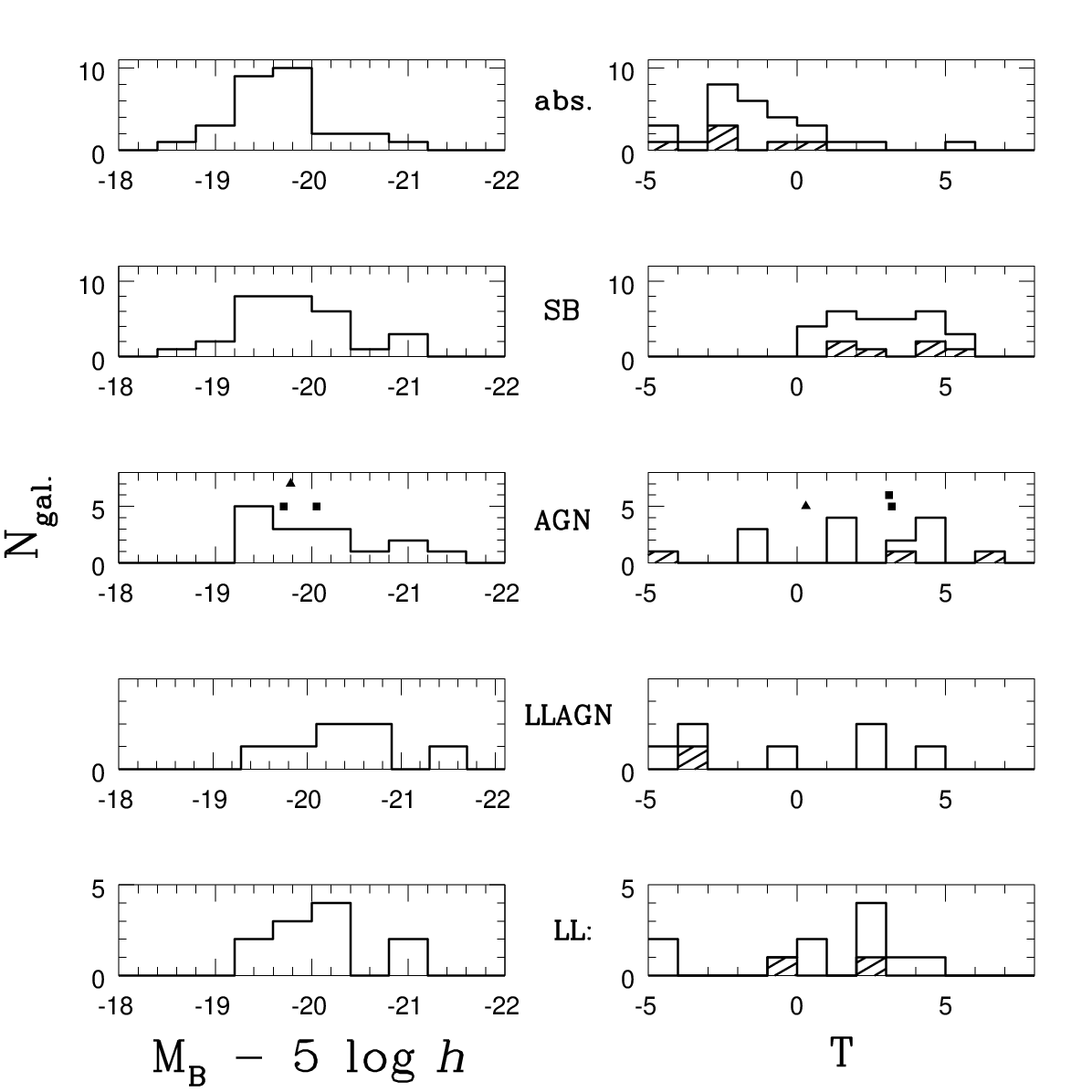}
  \caption{Luminosity (M$_B$) and morphology (T) distribution of absorption line galaxies, SBs,
  AGNs, LLAGNs and LLAGN candidates. The dashed distributions
  in the right panels indicate morphological distribution of the brightest 20 \% galaxies in each
  sample.In the AGN panels  are indicated the position of the 2 LINERs (squares) and the Sy 2
  (triangle) which have not been included in the histograms. }
    \end{figure}

Figure 8 shows the luminosity (M$_B$)  and morphology (T)
distribution of  absorption line galaxies. SBs, \,AGNs,\, LLAGNs
 and LLAGN candidates.
 M$_B$ has been derived from B$_T$  which is available in LEDA for all but one LLAGN galaxy (UZC-BGP 10B).
 The dashed histogram in the morphological distribution (right panels) refer to the 20 \%
 brightest galaxies in each sample. In AGN panels are also indicated magnitudes and types of
 the two LINERs (squares) and the Sy 2 (triangle) which have not been included in the histograms.
Examining the left panels of  Fig. 8 we see that galaxies displaying the largest B luminosity are LLAGNs,\,
in fact most (71 \%) of  them have M$_B$ $<$ -20 + 5 log $h$ a fraction which should be compared with 55 \% of
LLAGN candidates, 44 \% of AGNs (LINERs and Sy 2 included), 35 \%  of SBs and 18 \%  of absorption line galaxies.
These last ones are of particularly low B luminosity, quite unsual for passive galaxies which  are, in general,
much more luminous star forming ones (Kelm et al. 2005). Curiously, in our sample,
absorption line galaxies and SBs have rather similar luminosity distribution which derive from an almost
``opposite'' morphological content.
Further investigation is needed to understand
if and how this population of low luminous early-type passive galaxies is connected to pair environment or if it is a
more general
characteristics of low density large scale environment.

The right panels of Fig. 8 show that absorption line galaxies
are dominated by S0s (-3 $\le$ T $<$ 0) and that very few of these galaxies display nuclear activity,
ellipticals
(T $<$ -3), instead, show a much higher rate of nuclear activity and in some cases (AGNs and LLAGNs) are even among
the 20 \% brightest
galaxies in each sample. Thus, low B luminosity coupled to absence of emission features
seems to characterize more S0s than  ellipticals, which, however, are much rarer than S0s in our sample.

The morphological distribution of SBs is rather flat and equally flat is the distribution of the brightest 6
among them.  The first luminosity ranked ( M$_B$ = -21.15 + 5 log $h$ ) SB is a late spiral (UZC-BGP 9B)
 and the second one ( M$_B$ = -20.82 + 5 log $h$ ) UZC-BGP 2A, is an early spiral one.
The median type of SB morphological distribution is 2.95 (corresponding to Sb) while, Ho et al. (1997)
find a median value of 5 (corresponding to Sc) for their sample of galaxies with nuclear SB activity.
Our sample is surely early spiral dominated but we could get, however, a larger median value of T if we had
a larger content of late spirals among SBs. In our sample the fractions of early and late spiral
SBs are extremely similar (19/44 and 9/20), while Ho et al. (1997) report fractions  of 38 \% and 82 \%.
However, if we separate 4-lines SBs
(i.e. SB classified on the basis of at least 4 emission lines) from 2-lines SBs (classified
based on H$_{\alpha}$  and [NII]$_{6583}$ features) we find both a  B luminosity increase
(63 \% of these galaxies has  M$_B$ $<$ -20 + 5 log $h$)  and a more advanced morphological type (median T = 3.95).

The morphological distribution of AGNs is rather advanced, the median value is 2.55 ($\sim$ Sab)
to be compared with 1 ($\sim$ Sa) of Ho et al. (1997). AGNs are spread all over the morphological
whole range and the 3 brightest AGNs reflect this behaviour too. The brightest
(M$_B$ $=$ -21.6+ 5 log $h$) AGN is UZC-BGP 79A an elliptical galaxy
  (NGC 6251) classified as Sy 2
 (NED) showing, in our spectrum, an H$_{\beta}$  emission well below our adopted S/N threshold,
(see also sect. 3). The other two brightest
 galaxies have both M$_B$ $<$ -20.7+ 5
 log $h$ and advanced morphological type. The LINERs and the Sy 2, whose positions are
 represented
 with two squares and a triangle, display a modest luminosity. The morphological type of
 the LINERs is quite advanced too, for comparison Ho et al (1997) give a median value of 1.

LLAGNs and LLAGN candidates  distribute over the whole morphological range, their T morphological and luminosity
distributions are somewhat different,
LLAGN candidates displaying both a larger content of faint galaxies and spirals.

\begin{figure}
   \centering
  \includegraphics[width=8cm]{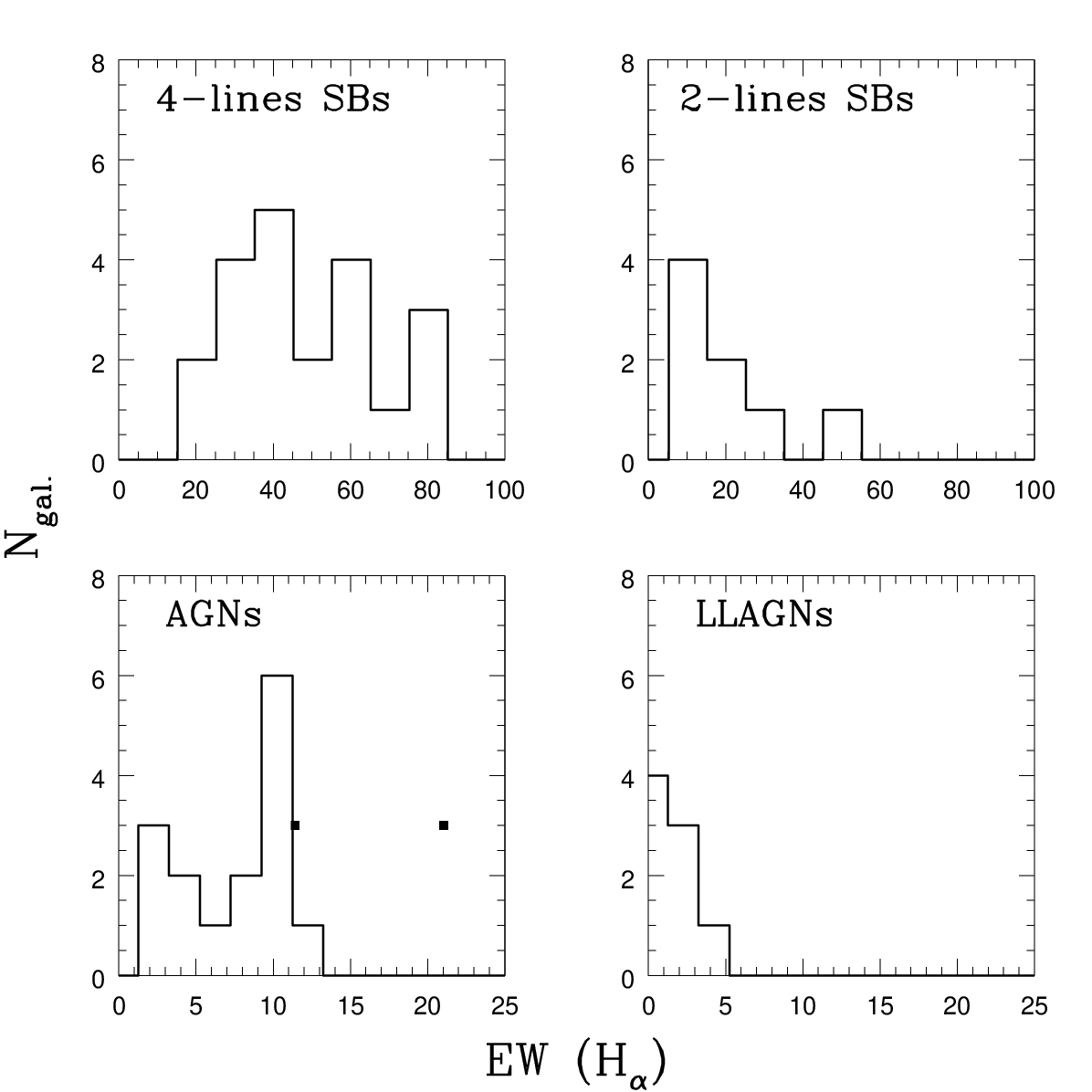}
  \caption{ H$_{\alpha}$ EW distribution of SBs classified on the basis of 4 (or more) emission lines
(upper left) and  of 2 emission lines (upper right), AGNs (lower left) and LLAGNs (lower right).
The squares in the AGN panel indicate the position which would be occupied by the two LINERs
(not included in the histogram). The Sy 2 position is not indicated as its H$_{\alpha}$ EW is
 the largest in the sample (117.43 \AA) and would fall well behind the
plot limits.}
    \end{figure}

 The EW of the H$_{\alpha}$ emission line of galaxies classified as SB (upper panels),
AGN (lower left)  and LLAGNs (lower right) is shown in Fig. 9. The upper left and right panels refer
respectively to 4-lines SBs and 2-lines SBs. The squares in the AGN panel indicate the position
which would be  occupied by the two LINERs, not included in the histogram. \,The position of the Sy 2
is not marked as it would be too much far on the right in this plot, this galaxy has, in fact, the largest (117.43 \AA)
 H$_{\alpha}$ EW  of  the whole sample.

The median values of the distribution of H$_{\alpha}$ EW of 4-lines
SBs, 2-lines SBs, AGNs and LLAGNs are respectively 45.17 \AA ,\,
16.48 \AA,\, 8.77 \AA \,(9.67 \AA \,including the Sy 2 and the
LINERs)  and 1.33  \AA.  \,For comparison Miller et al. (2003)
report, for a large sample of SDSS galaxies  having luminosity
similar to ours but belonging to any kind of environments, median
values of the H$_{\alpha}$\, EW of 26 \AA, \, 14 \AA \,and 3 \AA \,
for 4-lines SBs, 2-lines SBs and AGNs  respectively. \, Thus, while
the 2-lines SBs display similar values of the median H$_{\alpha}$
EW, both 4-lines SBs and 2-lines AGNs show larger values in ours
than in Miller et al. (2003) sample.\, However, the criterion
adopted by Miller et al. (2003) to classify 2-lines AGNs is more
conservative than the one adopted by us, as they required Log
([NII]/H$_{\alpha}$) $>$ -0.2. Application of  their criterion to
our data causes exclusion of 6 2-lines AGNs (belonging, obviously,
to the  large values tail of  H$_{\alpha}$ EW) and reduces the
median value of 2-lines AGNs, in our sample, to 4.94  a value
much more similar  but still exceeding (1.6 times) the one reported
by Miller et al. (2003).  The excess, however, fades completely 
out if we include LLAGNs in our AGN sample as the median value of
H$_{\alpha}$ EW drops to 2.34 \AA , which is even below the value 
(3 \AA) reported by Miller et al. (2003).
The inclusion of LLAGNs, among AGNs, is justified since Miller et al (2003) state that
a signficant population of their 2-lines AGNs are LLAGNs, although they
do not say how many LLAGNs enter their AGN sample. Since LLAGNs are characterized by lower 
values of H$_{\alpha}$ EW than AGNs, the H$_{\alpha}$ EW  of an AGN sample including also 
LLAGNs will reflect the relative amount of the two populations
entering the sample. For this reason we do not expect exact coincidence between 
the value that we derive in our sample and the one reported by Miller et al. (2003)
and we consider the agreement, between the two values, quite good.
Thus only 4-lines SB in our sample display a median  H$_{\alpha}$ EW
which is significantly larger than the one reported by Miller et al. (2003) 
and which can not be attributed to instrumental effects as we find no correlation between
the measured  H$_{\alpha}$ EW and the S/N of the spectra. 

In star forming galaxies the increase of  H$_{\alpha}$ EW relates to the Star Formation Rate (SFR,
Kennicutt \& Kent 1983; Kennicutt et al. 1987). Thus the larger  values displayed by  4-lines SB galaxies in our
 sample
compared to the ones of Miller et al. (2003) supports the interaction - SB scenario.
In this framework Barton et al. (2000) found a significant increase of H$_{\alpha}$ EW with decreasing
galaxy-galaxy projected distance
for SB galaxies in a magnitude limited sample of galaxy pairs. The correlation holds
on a limited projected distance range (from 5$h$ $^{-1}$ to 40 $h$ $^{-1}$ kpc) over which the
H$_{\alpha}$ EW decreases from about 150 \AA \, to 50 \AA.\,
Figure 9 shows that  there are no SBs in our sample with H$_{\alpha}$ EW larger than 85 \AA, that value
is reached only by one galaxy (UZC-BGP 83B) whose projected distance from its companion (the only Sy 2 that we
have detected in our sample) is 19 $h$$^{-1}$ kpc, which is small but not the smallest in the sample.
The two SBs showing the smallest projected distance (3  $h$$^{-1}$ kpc) in our sample
 belong to the same pair (UZC-BGP 74) and show
H$_{\alpha}$ EW of 58.94 and 57.30 \AA \, respectively.
 Half (15/29) of the SB in our sample display a value of
 H$_{\alpha}$ EW $>$ 40 \AA, which is supposed (Kennicut \& Kent 1984) to separate normal from intense SB activity.
None of these intense-SB galaxies is found in pairs with a projected galaxy
separation (r$_p$) larger than 160 $ h$ $^{-1}$
and 3 of them show r$_p$ $<$ 30  $h$ $^{-1}$. There are no other SBs with  r$_p$ $<$ 30 $ h$$^{-1}$, 
while for r$_p$ $>$ 160 $h$$^{-1}$
only 2 SBs are found (with an average value of H$_{\alpha}$ EW of 21.89 \AA).
We can not confirm the anticorrelation between  H$_{\alpha}$ EW and
galaxy-galaxy projected found by Barton et al. (2000), which is not
surprising since our sample is quite  different from theirs and we
have few (9) galaxy pairs characterized by small ($\sim$ 50$h$
$^{-1}$ kpc) galaxy-galaxy separation. The lack of a clear
anticorrelation in our sample might as well indicate that for bright
isolated galaxies
 in pairs interaction is at work and effective up to 
  160 $h$$^{-1}$ kpc.

\begin{figure}
   \centering
  \includegraphics[width=8cm]{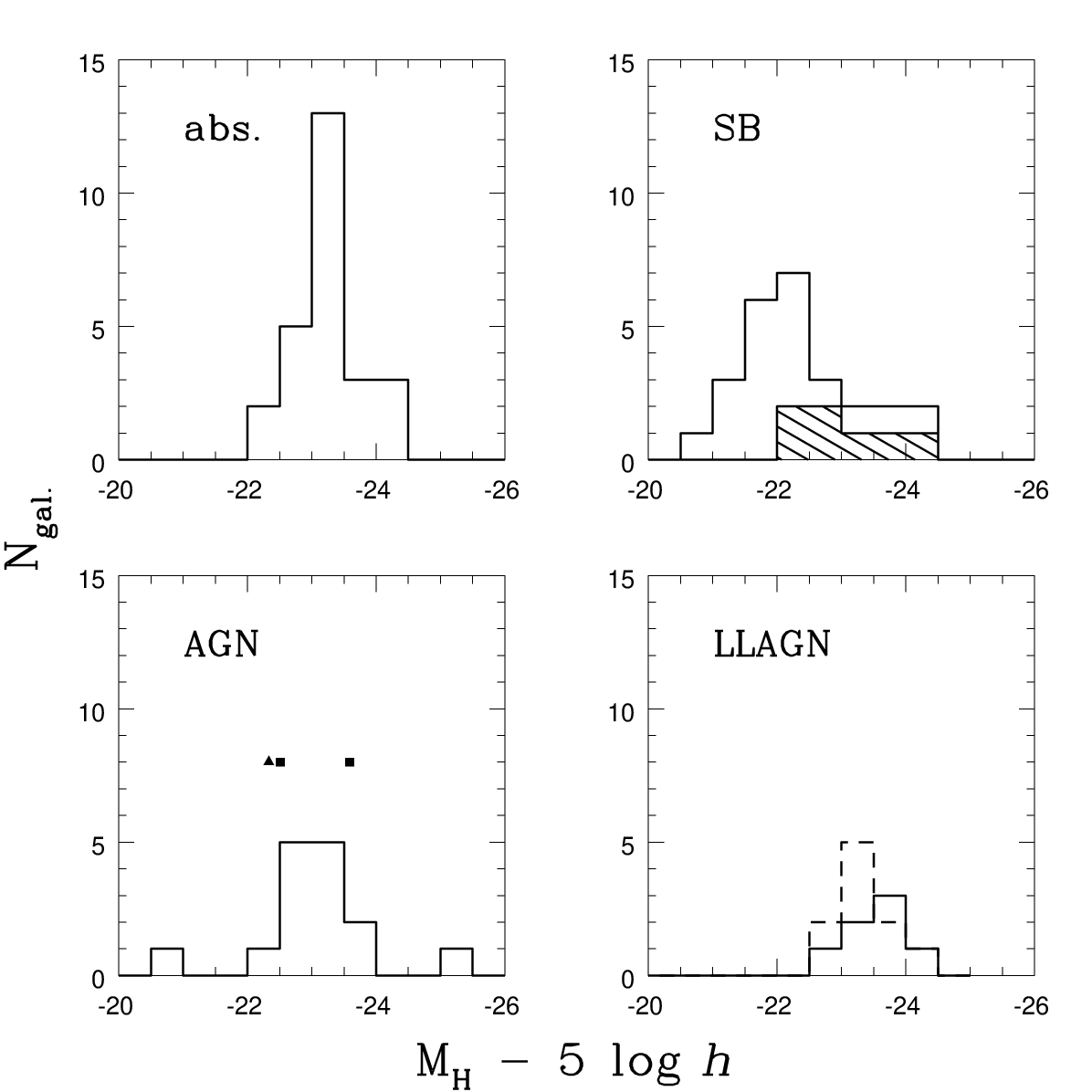}
  \caption{ M$_H$  distribution of galaxies with absorption spectrum (upper left)
and classified as  SB (upper right), AGN (lower left) and LLAGN (lower right). The continuous distribution
in the upper right panel represent the whole SB population, the dashed histogram 2-lines SBs.
In analogy with Fig. 8 we have marked in the AGN panel (lower left)
the position of the 2 LINERs (squares) and the Sy 2 (triangle) which have not been included in the histogram.
 In the lower right panel the continuous and dashed line refer respectively to  LLAGNs and LLAGN candidates.}
    \end{figure}

Figure 10 shows the M$_H$ distribution of galaxies in our sample
showing absorption lines (upper left) and classified as SB (upper
right), AGN (lower left) and LLAGN/LLAGN candidates (lower right,
continuous/dashed histogram). The  continuous distribution in the
upper right panel refers to the whole SB population, the dashed
histogram to 2-lines SBs. M$_H$ has been derived from 2MASS data
available for all but 6 galaxies in our sample, which are 3 SBs
(UZC-BGP 29B 49A and 74B), 2 absorption line galaxies (UZC-BGP 49B
and 58B) and 1 LLAGN (UZC-BGP  26B). At variance with Fig. 8 in
which galaxies  display rather similarly large B luminosity
distributions, Fig. 10 shows that only SBs have a broad M$_H$
distribution extending over 4 magnitudes, in all the other cases
galaxies are concentrated in a 2 magnitudes range. The only
exception are AGNs whose distribution, however, appear large due to
the presence of two galaxies only: the faintest and the brightest,
in H, of  the whole sample. The faintest  (M$_H$ =  -20.91 + 5 log
$h$) is  UZC-BGP 24B , an Sbc galaxy, the brightest (M$_H$ =  -25.03
+ 5 log $h$)
 is  UZC-BGP 79A an elliptical, whose Sy 2 nature we do not confirm due to our imposed emission line S/N threshold. \,
From Fig. 10 we see that while 10/26 SB galaxies have  M$_H$ $\ge$ -22 + 5 log $h$, there are no absorption line
galaxies fainter than that limit and only 1/18 AGNs.  LLAGNs and LLAGN candidates are even
brighter as they all show    M$_H$ $<$ -22.5 + 5 log $h$ .
The upper right panel of Fig. 10 shows that 2-lines SBs (dashed histogram) are all found at large H luminosity and
 constitute
more than half of the whole ``H luminous'' ( M$_H$ $>$  -22.5 + 5 log $h$) SB population. A KS test confirms (99.3 c.l.)
the difference between 4-lines SB and 2-lines SB M$_H$ distribution.
 The M$_H$  distribution of SBs, as a whole, differs significantly from the one of absorption line
galaxies, AGNs, LLAGNs and LLAGN candidates (KS c.l. $>>$ 99.9 \%,  99.7 \% , 99.1 \% and 99.9 \% respectively).
 The significance
of these differences becomes, obviously, larger if we exclude from
SB class the 2-lines SBs, (in this case we obtain a KS c.l. always
$>$ 99.9 \% for 4-lines SB vs. AGN, LLAGN and LLAGN candidates
respectively).

\begin{table*}
\begin{center}
\caption[] {The mass of galaxies with different nuclear activity type }
\begin{tabular}{||l|r|r|r|c|c|r|c|c||}
\hline
\hline
Activity  & N$_T$ & N$_H$ & N$_M$$_{dyn}$   & $\Delta$ {\it M}$_{dyn}$  & Med ($M_{dyn}$) & N$_{spir.}$
  & $\Delta$ {\it M}$_{spir.}$  & Med ($M_{spir.}$)\\
 & & & & (10$^{11}$ $M_{\sun}$) & (10$^{11}$ $M_{\sun}$) & & (10$^{11}$ $M_{\sun}$) & (10$^{11}$ $M_{\sun}$) \\
\hline
\hline
none  & 28  & 26 & 23  &  [1.0 : 5.5]  & 1.9 & 8 & [1.0 : 3.1] &  \\
SB    & 29  & 26 & 26 &  [0.3 : 4.5]  &0.8 &     &    &    \\
AGN   & 18  & 18 & 17 &  [0.2 : 3.8]  &1.5 & 15  &    &    \\
LLAGN & 8   & 7  & 5 & [1.2 : 3.2]  &  2.1 & 4  &     & 2.6 \\
LL:   & 10  & 10 & 8 & [1.1 :  6.0]   & 2.4 & 7 &     & 2.2 \\

\hline
\end{tabular}
\end{center}
\end{table*}

The luminosity in H relates to the galaxy mass within the optical radius of disc galaxies ($M_{dyn}$)
a quantity which can be derived through the relationship log ($M_{dyn}$/$M_{\sun}$) = log (L$_H$/L$_{H\sun}$) + 0.66
(Gavazzi et al. 1996).
Adopting  M$_H$ sun =3.39 (Allen 1973) we have calculated  $M_{dyn}$ for all the disc galaxies (i.e. elliptical
excluded)
in the sample having H magnitude available.
The results  are summarized in Table 7 where we indicate for each
activity type (column 1), the total number of galaxies (N$_T$)
(column 2), the number of galaxies (N$_H$) having H magnitude
available (column 3) and  the number of galaxies (N$_{M}$$_{dyn}$)
for which we have derived the galaxy mass ($M_{dyn}$), (column 4).
In column 5 we give the minimum and maximum value of the mass for
each distribution ($\Delta$ {\it M}$_{dyn}$) and in column 5 the median
value of each distribution (med ($M_{dyn}$)). With the term none
(first row of column 1) we indicate galaxies with absorption lines
only in their spectrum. Since the nature of S0 galaxies is somewhat
questioned and since among them there might be some misclassified
ellipticals we have also computed $M_{dyn}$ for spiral galaxies
only. The results, only if  different from values reported in
columns 3, 4 and 5,  are shown in columns 6, 7 and 8 in which we
indicate number of spiral galaxies (N$_{spir.}$), minimum and
maximum value
 of the mass ($\Delta$ {\it M}$_{spir}$) and median value of the mass distribution (med
($M_{spir}$)).
The  exclusion of S0s does not change significantly the results as
mass ranges remain the same, in all but the sample of absorption
line galaxies, and median values display, consequently, modest
changes (only for LLAGNs and LLAGN candidates).

Absorption line galaxies, LLAGNs and LLAGN candidates display the higher value of the mass, SB the lowest as this is the
only population with a median value of the mass distribution below 10$^{11}$ $M_{\sun}$. However if we
separate 2-lines SBs from 4-lines SBs the median value of the first ones increases to  1.3 10$^{11}$ $M_{\sun}$
and the mass range narrows (between 7.8  10$^{10}$ and 4.4 10$^{11}$ $M_{\sun}$), while the second ones
display a mass range between 2.5  10$^{10}$ and  4.5 10$^{11}$ $M_{\sun}$  and  median value of  6.5  10$^{10}$   $M_{\sun}$.
AGNs appear characterized by a wide mass distribution  too, which is induced by the presence of an extremely
low mass galaxy (UZC-BGP 24B, see also Fig.10 and relative comments). Exclusion of this galaxy from the AGN
sample would raise  the minimum mass  to 7.9  10$^{10}$ (still below the minimum for absorption line
galaxies, LLAGNs and LLAGN candidates)  and the median value to  1.6 10$^{11}$ $M_{\sun}$.

Kauffmann et al. (2003) have shown that AGNs reside only in galaxies with masses $M$ $>$ 10$^{10}$ $M_{\sun}$,
a value that should be scaled to 0.7 10$^{10}$ $M_{\sun}$ to account for the different value of H$_0$ adopted by us
and by them. Table 7 shows that, due to the imposed criterion on luminosity (see also sect. 2), 
all galaxies 
in our sample 
have masses  above the Kauffmann et al. (2003) value, we are thus unable to confirm
their finding. 

\begin{figure}
   \centering
  \includegraphics[width=8cm]{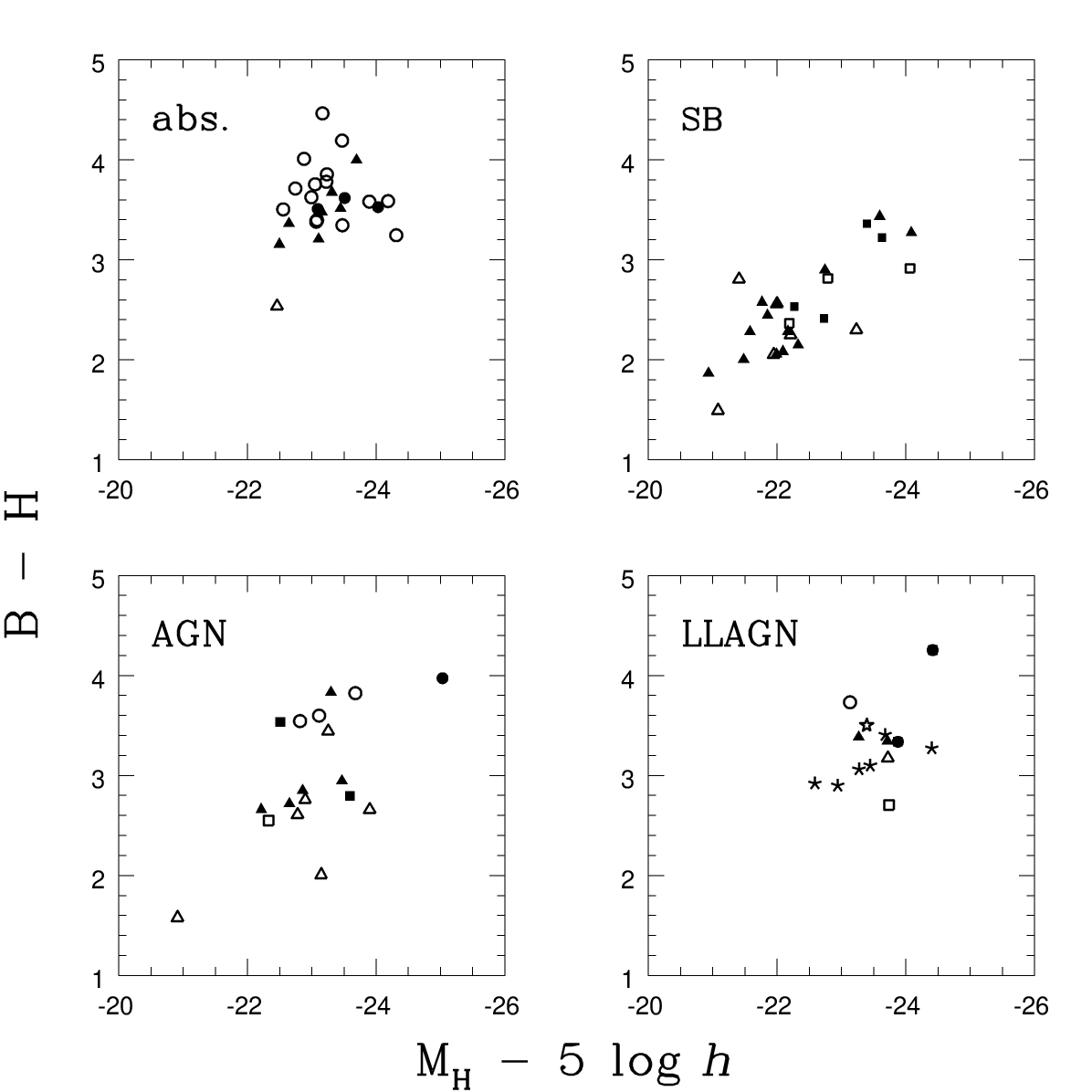}
  \caption{Color magnitude diagram (B-H vs. M$_H$) for galaxies with absorption spectra (upper left)
and classified as SBs (upper right), AGNs (lower left), LLAGNs and LLAGN candidates (lower right).
Filled and empty circles and triangles indicate, in each panel ellipticals, S0s, early and late spirals.
In the SB panel (upper right) the previously defined symbols indicate morphological classification of
4-lines SBs, while filled and empty squares indicate early and late
spirals of 2-lines SBs. In the AGN panel (lower left) filled and empty squares indicate,
instead, the 3 galaxies classified respectively as LINERs  and Sy 2. In the LLAGNs/LLAGN candidates panel
(lower right)
the standard symbols (filled/open circles/triangles) have been used to distinguish LLAGN morphological types
while we have used filled/open squares to indicate LLAGN candidates ellipticals and S0s and  filled/empty
stars to indicate early/late spirals. }
    \end{figure}

Figure 11 illustrates the Color Magnitude (CM) diagram (B-H vs. M$_H$)
 for galaxies with absorption spectra (upper left) and classified as SB (upper right),\,
AGN (lower left), \,LLAGN and LLAGN candidates (lower right).
We distinguish, for each activity class, different morphological types. As a general rule
filled and empty circles indicate ellipticals and S0s, while filled and empty triangles stand for early and late spirals.
This rule holds for the absorption line galaxies, 4-lines SBs (although this class of objects
does not contain early-type galaxies), AGNs  and LLAGNs.
Concerning 2-lines SBs (upper right panel) early and late spirals  are indicated as filled and empty squares,
while for LLAGN candidates (lower right panel) we have adopted filled and empty squares to indicate E and S0s
and filled and empty stars  to indicate early and late spirals.
Finally in the AGN panel (lower left)
empty and filled squares indicate the Sy 2 and LINERs location respectively (all three are early spirals).

Absorption line galaxies (upper left panel of Fig. 11) occupy a well
defined ``red-bright'' region in the CM diagram, with the exception
of  UZC-BGP 8A (NGC 800) which is the only late type spiral, with an
absorption spectrum,  in our sample. The absence of a trend for
absorption line galaxies in the CM diagram must be attributed to the
large group of S0s (empty circles) which dominate this population.
In fact early spirals with an absorption spectrum (filled triangles)
follow the expected trend and get more H luminous as they become
redder (B-H increasing). SB galaxies (upper right panel) follow a
much better defined sequence which holds on a wide range of
luminosities. The large proportion of filled vs. empty symbols
indicate the dominance of early spirals among this class of objects.
The faint luminosity region is mostly occupied by a population
of early spirals which besides being quite faint  (M$_H$ $\leq$ -22 + 5 log $h$) are also exceptionally
``blue'' (B-H $\sim $ 2.2 ).
As expected (cfr. Fig. 9)  early and late spirals classified as 2-lines SBs (filled and empty squares) occupy the high H luminosity
 ''red'' region of the diagram. They too follow the general SB trend. The AGN panel shows the absence of a
trend and all but 2 (the brightest and the faintest) AGNs are concentrated in a narrow
region  which resembles the absorption line galaxies ``cloud'' but which is characterized by a bluer color
and a somewhat fainter H luminosity. The  AGN ``cloud'' is  actually made of two different regions
which are characterized by similar H luminosity and different color.  The separation of the two AGN regions
occurs at B-H $\sim$  3, which is also the value below which almost no absorption line galaxies are found
and where most SB galaxies reside. Also LLAGN and all but one LLAGN candidates
occupy a well defined region in the CM diagram which is brighter than the ones occupied by absorption line
galaxies and AGNs and has the color in between the two.
 As a whole  Fig. 11 indicates that absorption line galaxies, SBs, AGNs and LLAGNs
occupy rather distinct regions in this CM diagram so that it would  possible to
distinguish SB from absorption line galaxies only on the basis of their location in this diagram.
AGN location is in between the previous two, but almost completely separated from the one of LLAGN and LLAGN candidates.
This is an interesting finding that should be checked on galaxy samples in different environments.

 \section{ Nuclear activity in pairs showing evident interaction }

There are cases in which galaxy interaction  produces
``visible'' effects such as morphological distortions, bridges and
tidal tails. The detection of those features  depends, however, on
the sensitivity to low surface brightness structures and is thus
strongly affected by seeing, depth and redshift. Moreover
morphological distortions are short living phenomena (100 Myr) and
tidal tails appear only in prograde disk encounters (Toomre \&
Toomre 1972). Thus a selection of systems based on the presence of
''visible'' signs of interaction would surely lead to miss a
fraction of interacting pairs. However, pairs showing evident
interaction can be used to test the pair sample as a whole, as
differences that might be present in the content and type of nuclear
activity between these systems and the whole pair sample could bring
into question the interaction status of galaxy pairs.

Careful inspection of DSS images of all pairs in our sample allowed
us to identify 9 pairs in which interaction is clearly ``visible''
between the members. These pairs are listed in Table 8, where we
give, for each of them, identificator (column 1), pair/galaxy
morphology  (column 2; as in Tables 4, 5 and 6 galaxy morphology of
the active member is underlined in case of E+S pair), nuclear
 activity type (column 3; -- indicate absence of nuclear activity)
and  H$_{\alpha }$ EW
(column 4; only if the galaxy is active).

\begin{table*}
\begin{center}
\caption[] {UZC-BGPs showing interaction between members}
\begin{tabular}{||r|c|l|l|r|r|r||}
\hline
\hline
UZC-BGP & Pair type & \multicolumn{2}{c|}{Activity type} & \multicolumn{2}{c|}{H$_{\alpha}$ EW} \\
\cline{3-6}
& & A & B & A & B  \\
\hline
\hline
8 & S+S & -- & SB & -- & 12.46  \\ 
14 & E+$\underline{S}$ & SB & -- & 74.51 & --  \\ 
21 &E+$\underline{S}$& -- & LLAGN &-- &  0.31 \\ 
49 & S+S & SB & --  & 20.51 & --  \\ 
59 & S+S & LL: &AGN &  -- & 9.39 \\ 
65 & E+$\underline{S}$ & SB & -- & 43.47 & -- \\ 
71 & S+S & LL: & AGN   &-- & 1.84 \\ 
74 & S+S & SB & SB & 58.94 & 57.30 \\ 
83 & S+S& Sy2 & SB  & 117.43  & 85.01 \\ 
\hline
\end{tabular}
\end{center}
\end{table*}



\begin{table*}
\begin{center}
\caption[] {UZC-BGP galaxies interacting  with fainter companions }
\begin{tabular}{||r|c|l|r|l||}
\hline
\hline
UZC-BGP & Pair type & Activity type & H$_{\alpha}$ EW & Companion \\ 
\hline
\hline
4B & E+$\underline{S}$& LLAGN & 0.34 & IC 1559  \\ 
5B & $\underline{E}$+S& -- & -- & NGC 197  \\
9B & S+S &SB & 11.22  & NGC 876  \\ 
10A &$\underline{E}$+S &LLAGN & 1.66 & NGC 997N \\ 
16A & $\underline{E}$+S& -- & -- & NGC 1588 \\ 
20A & E+$\underline{S}$  &SB & 76.86 & NGC 2744S \\ 
20B & $\underline{E}$+S & LLAGN & 1.02 & NGC 2751 \\ 
26B &$\underline{E}$+S &LLAGN & 0.47 & IC 590 E   \\ 
28A &S+S &LINER & 11.42 & NGC 3303 N \\ 
\hline
\end{tabular}
\end{center}
\end{table*}

\begin{table*}
\begin{center}
\caption[] {UZC-BGPs with absorption line spectrum in both members }
\begin{tabular}{||r|c|l||}
\hline
\hline
UZC-BGP & Pair type & Interaction \\
\hline
\hline
23 & E+E & A with NGC 2988  \\ 
31 & S+S & no\\ 
35 & E+E &  A with B \\
58 & E+E & A with B \\ 
63 & E+E & no \\ 
67 & E+S & no \\ 
76 & S+S & no \\ 
\hline
\end{tabular}
\end{center}
\end{table*}

Among the 48 galaxy pairs which constitute our sample only 9 (listed in Table 8) are characterized by  evident
 signs of interaction between
their members. Table 8 shows that this happens  more
frequently in S+S (6/23) than in E+S (3/19) pairs and that in these
last cases interaction is never able to induce emission activity in the early
type member of the pair. On the other hand  evident  interaction does
not appear to be able to induce nuclear activity in 2 galaxies
belonging to the S+S pairs  either. All kinds of activities are found in
pairs listed in Table 7;  visible interaction thus does not appear to
be responsible for activation of a particular kind of nuclear
activity. In Table 8 there are 7 SB galaxies, 3 AGNs (Sy 2 included), 1 LLAGN , 2 LLAGN candidates
and 5 galaxies with absorption spectrum. This implies fractions of activity type per  galaxy of 39 \%, 17 \%,
6 \%, 11 \% and 28 \% which are remarkably similar to the ones holding
for the whole sample (30 \%, 19 \%, 8 \%, 11 \% and 30 \%). Five of the SBs in Table 8
are 4-lines SBs, two (UZC-BGP 8A and 49A) are 2-lines SBs. Thus also the fraction of 2 over 4-lines SBs
in these pairs agrees remarkably well with the fraction in the whole sample (8/21). From column 5
of Table 8 we can derive the median value of H$_{\alpha}$ EW  of 4-lines and 2-lines SB respectively
58.94 \AA \, and 16.49 \AA ,\,again very similar to the ones derived for the whole sample (cfr. sect. 4).
The similarity in terms of amount, type and characteristics of the nuclear activity content between
these 9 pairs and the whole sample
supports the hypothesis that galaxy-galaxy  interaction between bright galaxies
in isolated pairs is at work and effective .

Since UZC-BGP is a volume-limited sample, evident interaction may
even occur between one pair member and a faint companion which  has
gone undetected either by UZC catalog magnitude limit (m$_{Zw}$
$\le$ 15.5) or by UZC-BGP luminosity limit (M$_{Zw}$ $\le$ -18.9 + 5
log $h$). This happens for 9 galaxies, belonging to 8 distinct
pairs, which we list in Table 9. For each galaxy we give
identificator
(column 1), pair/galaxy morphology (column 2; as in Table 8 galaxy morphology is underlined in case of E+S ),  activity type (column 3;
as in Table 8  --  means no activity), H$_{\alpha}$ EW, identificator
of the companion (column 5).
Among these faint companions only two (NGC 1588 and  NGC 2751)
interacting respectively with UZC-BGP 16A and 20B have failed
the luminosity limit of UZC-BGP, in all other 7 cases companions were too faint
to be included in the UZC catalog.


At variance with Table 8, Table 9 shows that most (7/9) of the galaxies
interacting with fainter companions belong to E+S pairs.
Five galaxies in Table 9 are early-type, two of them
(UZC-BGP 5B and 16A) display no activity while the remaining three (UZC-BGP 10A, 20B and 26B)
show LLAGN activity. This last one is the most frequent kind of activity of galaxies in
 Table 9. The fraction of LLAGNs among the galaxies of Table 9 is 44 \% (to be compared
with 8 \% on the whole sample). The excess of LLAGNs could indicate minor
galaxy interaction (Woods \& Geller 2007) as a driving mechanism for this kind of activity.
Interestingly the interacting LLAGNs of Table 9 are all found in E+S pairs which we expected (cfr. Fig.3 and discussion)
to be possibly, in large part, the bright core of a looser structure. The presence of a fainter close companion
confirms both our expectation and previous finding (Coziol et al 1998, 2000; Martinez et al. 2006) which claim
LLAGN to be the most common kind of activity in Compact
Groups.
Further investigation is needed since no detailed analysis concerning
the link between LLAGN activity and interaction has been carried out so far.


Finally, if interaction plays a major role in activating nuclear
activity passive galaxies in galaxy pairs should not display evident
interaction signs. There are 7 pairs in our sample with both members
passive. They  are listed in Table 10 where we give, pair
identificator (column 1), morphology (column 2) and, in column 3,
we indicate if interaction occurs between pair members or with a
faint companion. Only one galaxy (UZC-BGP 23A) interacts with a
faint companion (NGC 2988 which was not included in UZC catalog).
The sample is dominated by (4/7) by E+E pairs , 3 of which show
interaction patterns, the remaining 2 S+S and E+S pairs do not show
interaction at all. There is evident interaction and no activity
only in the 3 E+E pairs, which is somewhat encouraging for the
interaction-activity scenario as galaxy interaction is surely  going
to produce larger effects in gas rich than in gas poor galaxies,
while the 2 S+S pairs are characterized by large values ( 159 and
183  $h$$^{-1}$ kpc) of r$_p$ (cfr. Fig. 3, upper middle panel) in
our sample. This confirms our previous finding on H$_{\alpha}$ EW
and suggest that in UZC-BGP sample interaction is at work and
effective up to 160 $h$$^{-1}$ kpc.

 \section{Conclusions}

 To investigate the role of galaxy interaction on nuclear activity
 we have performed a detailed spectroscopical analysis on
 48 galaxy pairs, which represent more than half of the whole UZC-BGP sample
 and have an excellent morphological match with it. 

 We have found an extremely large fraction of emission line galaxies
 in our sample particularly among early (84 \%) and late (95 \%) spirals.

 Classification and analysis of spectral activity, performed by means of standard diagnostic diagrams,
 allowed us to show that  SB is the most frequent (30 \% of galaxies) kind of activity in our sample.
 It occurs exclusively in spiral galaxies with a frequency of SB phenomenon
 among spirals of 45 \%. The blue luminosity distribution of these SB is not particularly
 high as 67 \% have M$_B$ $>$ -20 +5log$h$.

 While SB are preferentially found in S+S pairs, AGN are almost equally found in S+S and in E+S pairs
 although in most cases (82\%) this kind of activity is displayed by a spiral galaxy. AGN
 in our sample show, in fact,a rather advanced
 morphological distribution characterized by a high blue luminosity. The fraction of AGNs
 in late spirals is 35 \% and late spirals hosting AGNs have an average M$_B$ = -20.4 + 5log$h$.

SBs display enhanced H$_{\alpha }$ EW an effect which relates to star formation and
might thus be related to pair environment. Star formation turns out 
to be intense in half of the SB
galaxies in our sample. Intense-SBs  have galaxy-galaxy separations up to 160 $h$$^{-1}$ kpc
 implying that interaction may be effective in isolated pairs of bright galaxies up to that distance.

Absorption line galaxies, SBs, AGNs and LLAGNs (candidates included) occupy rather distinct locations
in the B-H vs M$_H$ diagram a characteristics which reflects the different distribution
in B and H luminosity of each sample. Galaxy masses, estimated using the H luminosity, are
high for absorption line galaxies and LLAGNs (as a whole), low for SBs and "intermediate" for AGNs

All LLAGNs reside in E+S  and are equally
 distributed between early type galaxies and spirals. We have shown that half of them
 are hosted in galaxies displaying visible signs of interaction with fainter companions, which
 suggests that minor interaction might be a driving mechanism from a relevant fraction of
 LLAGNs. LLAGN has been claimed to be a heterogenous class of objects: our previous finding
concerning only half of the whole LLAGN population, coupled with the quite different behaviour in
terms of blue luminosity and morphological content of LLAGN candidates appear to  confirm that claim.

\begin{acknowledgements}
    This work was supported by MIUR. P.F acknowledges financial support
from the contract ASI-INAF I/023/05/0. S.M. acknowledges a fellowship by INAF-OAB.
    This research has made use of  the NASA/IPAC Extragalactic Database (NED)
    and of the Hyperleda Database (http://leda.univ-lyon1.fr/).
We thank an anonymous referee whose comments and criticism greatly improved the
scientific content of the paper.

\end{acknowledgements}


\begin{thebibliography}{}

\bibitem [Allen 1973]{Allen73}
Allen, C. 1973, Astrophysical Quantities, The Athlone Press (London)

\bibitem [Alonso et al. 2004] {Alonso04}
Alonso, M.S., Tissera, P.B., Colwell, G., \& Lambas, D. 2004, MNRAS, 352, 1081

\bibitem[Bahcall et al. 1995]{bahcall95}
Bahcall, N.A., Lubin, J.M., \& Dorman V. 1995, ApJ, 447, L81

 \bibitem [Baldwin, Phillips \& Terlevich 1981] {BPT1981}
     Baldwin, J. A., Phillips, M. M., \& Terlevich, R. 1981, PASP, 93, 5

  \bibitem[Barnes \& Hernquist 1991]{barnes91}
  Barnes J.E., \& Hernquist L.E. 1991, ApJ, 370, 65

  \bibitem[Barton 2000]{barton00}
  Barton, E.J., Geller M.J., \& Kenyon S.J. 2000, ApJ, 530, 660

  \bibitem[Borne et al. 1999]{borne99}
  Borne, K.D., Bushouse, H., Colina, L., et al., 1999, Ap \& SS, 266, 137


\bibitem[Bundy et al. 2000]{bundy00}
Bundy, K., Fukugita, M., Ellis, R.S., Kodama, T., Conselice, C.J., 2000, ApJ, 601, 123

 \bibitem[Carlberg et al. 1994]{carlberg94}
  Carlberg, R.G., Pritchet, C.J., \& Infante, L., 1994  ApJ 435, 540

 \bibitem[Coziol et al. 1998]{coziol98}
 Coziol, R., Ribeito, A.L.B., de Carvalho, R.R., \& Capelato, H.V. 1998, ApJ, 493, 563

\bibitem[Cid Fernandes et al. 2004]{fernandes04}
Cid Fernandes, R.,  Gonzales D., Rosa M., et al. 2004, ApJ 605, 105

\bibitem[Cuesta-Bolao \& Serna 2003]{cuesta03}
Cuesta-Bolao, M.J., \& Serna, A. 2003, A\&A, 405, 917

  \bibitem[Dahari 1985]{dahari85}
  Dahari, O. 1985, ApJS, 57, 643

  \bibitem[De Robertis et al. 1998]{derobertis98}
  De Robertis, M.M., Hayhoe, K., \& Yee, H.K.C. 1998, ApJS, 115, 163

   \bibitem[Donzelli \& Pastoriza 1997]{donzelli97}
   Donzelli, C. \& Pastoriza, M.G. 1997, ApJS, 111, 181

 \bibitem[Dubinski et al. 1996]{dubinski96}
Dubinski, J., Mihos, J.C., \& Hernquist, L., 1996, ApJ 462, 576

  \bibitem[Falco et al. 1999]{falco99}
   Falco, E.E., Kurtz, M.J., Geller, M.J., et al. 1999, PASP, 111, 438


  \bibitem[Focardi \& Kelm, 2002]{focardi02}
  Focardi, P., \& Kelm, B. 2002, A\&A, 391, 35

\bibitem[Focardi et al 2006]{focardi06}
Focardi, P., Zitelli, V., Marinoni, S., \& Kelm, B. 2006, A\&A, 456, 467 (paper I)

  \bibitem[Fuentes-Williams \& Stocke 1988]{fuentes88}
   Fuentes-Williams, T., \& Stocke, J.T. 1988, AJ, 96, 1235

\bibitem[Gavazzi et al. 1996]{gavazzi96}
Gavazzi, G., Pierini, D., \& Boselli, A. 1996, A\&A, 312, 397

    \bibitem[Gon\c{c}alves te al. 1999]{Goncalves1999}
     {Gon\c{c}alves, A. C., V\'{e}ron-Cetty, M.-P., \& V\'{e}ron, P.} 1999
     A\&AS, 135, 437

\bibitem[Gottlober et al. 2001]{Gottlober01}
Gottlober, S., Klypin, A., \&  Kravstov, A.V., 2001, ApJ 546, 223

\bibitem[Governato et al. 1999]{Governato99}
Governato, F., Gardner, J.P., Stadel, J., Quinn, T., \& Lake, G., 1999, MNRAS 307, 949


    \bibitem[Ho  1994]{Ho94}
   Ho, L.C., Filippenko, A.V., \& Sargent, W.L.W., 1994, IAU Symp., 159, 275

\bibitem[Ho  1997]{Ho97}
Ho, L.C., Filippenko, A.V., \& Sargent, W.L.W. 1997, ApJ, 487, 568

  \bibitem[Karachentsev 1972]{kara72}
   Karachentsev, I.D. 1972, in Catalogue of Isolated Pairs in the Northern Emisphere, Comm. Spec. Ap.
   Obs., 7, 1 (KPG)

 \bibitem[Kauffmann et al. 2003]{kauff03}
 Kauffmann. G.,Heckman, T.M., Tremonti, C., et al. 2003, MNRAS, 346, 1055

    \bibitem[Keel et al. 1985]{keel85}
   Keel, W.C., Kennicutt, R., Hummel, E., \& van der Hulst, J. 1985, AJ, 90, 708

   \bibitem[Keel  1993]{keel93}
   Keel, W.C. 1993, AJ, 106, 1771

   \bibitem[Keel  1996]{keel96}
   Keel, W.C. 1996, AJ, 111, 696

   \bibitem[Kelm et al.  1998]{kelm98}
   Kelm, B., Focardi, P.,  \& Palumbo, G.G.C. 1998, A\&A, 335, 912

   \bibitem[Kelm  et al.  2004]{kelm04}
   Kelm, B., Focardi, P., \& Zitelli V. 2004, A\&A, 418, 25

    \bibitem[Kelm \& Focardi  2004]{kelm04b}
   Kelm, B., \& Focardi, P.  2004, A\&A, 418, 937

   \bibitem[Kelm  et al.  2005]{kelm05}
   Kelm, B., \& Focardi, P., \& Sorrentino, G., 2005, A\&A, 442, 117

\bibitem [Kennicutt \& Kent 1983]{kennicut83}
 Kennicutt, R.C.Jr.,\& Kent, S.M., 1983, AJ, 88, 1094

   \bibitem[Kennicutt \& Keel 1984]{kennicut84}
    Kennicutt, R.C.Jr.,\& Keel, W.C. 1984, ApJ, 279, L5


       \bibitem[Kennicutt et al. 1987]{kennicut87}
       Kennicutt, R.C.Jr., Kennicutt, R.C.Jr., Keel, W.C., van der Hulst, J.M., Hummel, E., 1987, AJ, 93, 1011



\bibitem[kennicutt 1994]{kennicut94}
Kennicutt, R.C.Jr., Tamblin, P., \& Congdon, C.E. 1994, ApJ, 435, 22

          \bibitem[Kewley et al.  2001]{kewley01}
       Kewley, L., Dopita, M., Sutherland, R., Heisler, C., Trevena, J. 2001, ApJ, 556, 121

     \bibitem[Kewley et al.  2006]{kewley06}
    Kewley,L.J., Groves, B., Kauffmann, G., \& and Heckman, T. 2006, MNRAS, 372, 961


  \bibitem[Larson \& Tinsley 1988]{larson88}
    Larson, R.B., \& Tinsley, B.M. 1978, ApJ, 219, 46L

    \bibitem[Lauberts \& Valentijn 1989]{lau89}
    Lauberts, A.,  \& Valentijn, E. 1989, The Surface Photometry Catalogue of the ESO-Uppsala Galaxies,
    ESO, Garching bei Munchen (ESO-LV)

 \bibitem[Le Fevre et al. 2000]{lefevre00}
Le Fevre, O., Abraham, R., Lilly, S., 2000, MNRAS, 311, 565

    \bibitem[Mackenty 1989]{mckenty89}
    MacKenty, J.W. 1989, ApJ, 343, 125

\bibitem[Martinez et al. 2006]{martinez06}
Martinez, M.A., del Olmo, A., Perea, J., \& Coziol, R. 2006, in Groups of Galaxies in the
Nearby Universe, ESO-workshop, [astro-ph/0611098]

\bibitem[Mihos \& Hernquist 1996]{mihos96}
Mihos, J., \& Hernquist, L., 1996, ApJ 464, 641


\bibitem[Miller et al. 2003]{miller03}
Miller, C.J., Nichol, R.C., Gomez, P.L., Hopkins, A.M., \& Bernardi, M. 2003, ApJ, 597, 142

  \bibitem[Noguchi 1988]{noguchi88}
  Noguchi, M.1988, A\&A, 203, 259

 \bibitem[patton 2000]{patton00}
Patton, D.R., Carlberg, R.G., Marzke, R.O., et al., 2000, ApJ 536, 153

 \bibitem[paturel 2003]{paturel03}
  Paturel, G., Petit, C., Prugniel, Ph., et al. 2003, A\&A 412, 45

    \bibitem[Rafanelli et al. 1995]{rafanelli95}
  Rafanelli, P., Violato, M., \& Baruffolo, A. 1995, AJ, 109, 1546

\bibitem[Reduzzi \& Rampazzo 1995]{reduzzi95}
Reduzzi, L., \& Rampazzo, R. 1995, ApJL \& Comm., 30, 1 (RR)

\bibitem[Sanders \& Mirabel 1996]{sanders96}
    Sanders, D.B.,\& Mirabel, I.F. 1996, ARA\&A, 34, 749

\bibitem[Sanders et al. 1988]{sanders88}
    Sanders, D.B., Soifer, B.T., Elias, J.H., et al. 1988, ApJ, 325, 74

\bibitem[Schmitt 2001]{schmitt01}
    Schmitt, H.R. 2001, AJ, 122, 2243



 \bibitem [Toomre1972]{toomre72}
 Toomre, A., \& Toomre, J. 1972, ApJ, 178, 623

     \bibitem[Veilleux \& Osterbrock 1987]{VO1987}
     Veilleux, S., \& Osterbrock, D. E. 1987, ApJS, 63, 295

     \bibitem[Veilleux et al. 1995]{Veilleux1995}
     Veilleux, S., Kim, D.C., Sanders, D.B., Mazzarella, J.M., \& Soifer, B.T. 1995,
     ApJS, 98, 171

    \bibitem[Veilleux 2002]{Veilleux2002}
     Veilleux, S. 2002, in AGN Surveys, ed. R.F. Green, E.Ye. Khachikian, \& D.B.Sanders, IAU Coll. 184,
     ASP Conf. Proceed. 284, 111


    \bibitem[V\'{e}ron et al. 1997]{Veron1997}
     V\'{e}ron, P., Gon\c{c}alves, A.C., \& V\'{e}ron-Cetty, M.P. 1997
     A\&A, 319, 52


     \bibitem[Woods \& Geller 2007]{woods07}
     Woods, D. F., \& Geller, M. 2007, AJ 134, 527

\bibitem[Zepf \& Koo 1989]{Zepf1989}
     Zepf, S.E., \& Koo, D.C., 1989, ApJ 337, 34


\end{thebibliography}
\end{document}